\documentclass{article}
\usepackage[margin=1in]{geometry}
\usepackage{comment}
\usepackage{graphicx} 
\usepackage{subcaption}
\usepackage[utf8]{inputenc}
\usepackage[mathscr]{eucal}
\usepackage[all]{xy}
\usepackage{amsmath}
\usepackage[dvipsnames]{xcolor}
\usepackage{amsthm, amssymb, amscd, amsxtra,color,authblk,tikz,arydshln,array, caption, romannum,verbatim,enumitem}
\usepackage[section]{placeins}
\usetikzlibrary{decorations.markings}
\usepackage{booktabs}

\title{Floquet Codes from Coupled Spin Chains}
\author[1]{Bowen Yan $ ^*$}
\author[1]{Penghua Chen}
\author[1,2]{Shawn X. Cui \thanks{Corresponding author}}
\affil[1]{{\small Department of Physics and Astronomy, Purdue University, West Lafayette}}
\affil[2]{{\small Department of Mathematics, Purdue University, West Lafayette}}
\affil[ ]{{\small \it \{yan312,  cui177,chen3014\} @purdue.edu}}

\begin{document}
\pagenumbering{arabic}

\maketitle

\begin{abstract}

We propose a novel construction of the Floquet 3D toric code and Floquet $X$-cube code through the coupling of spin chains. This approach not only recovers the coupling layer construction on foliated lattices in three dimensions but also avoids the complexity of coupling layers in higher dimensions, offering a more localized and easily generalizable framework. Our method extends the Floquet 3D toric code to a broader class of lattices, aligning with its topological phase properties. Furthermore, we generalize the Floquet $X$-cube model to arbitrary manifolds, provided the lattice is locally cubic, consistent with its Fractonic phases. We also introduce a unified error-correction paradigm for Floquet codes by defining a subgroup, the Steady Stabilizer Group (SSG), of the Instantaneous Stabilizer Group (ISG),  emphasizing that not all terms in the ISG contribute to error correction, but only those terms that can be referred to at least twice before being removed from the ISG. We show that correctable Floquet codes naturally require the SSG to form a classical error-correcting code, and we present a simple 2-step Bacon-Shor Floquet code as an example, where SSG forms instantaneous repetition codes. Finally, our construction intrinsically supports the extension to $n$-dimensional Floquet $(n,1)$ toric codes and generalized $n$-dimensional Floquet $X$-cube codes.
\end{abstract}

\section{Introduction}

Kitaev proposed the paradigm for topological quantum computation via manipulating anyons \cite{kitaev_fault-tolerant_2003}. The information is stored in the fusion space of anyons, and quantum gates are accessed by braiding, thus not affectable at the microscopic level, making them naturally fault-tolerant to local perturbations. This paradigm opens new horizons for quantum computation, while the anyon theory itself, or the topological phase of matter that supports anyons, remains a topic of great theoretical interest. Among these, exactly solvable topological lattice models, such as the Kitaev quantum double model \cite{kitaev_fault-tolerant_2003}, the string-net model \cite{levin2005string}, and the Kitaev spin liquid model \cite{kitaev_anyons_2008}, are excellent examples that exhibit the typical nature of topological phases of matter.

Exactly solvable topological models are typically associated with frustration-free Hamiltonians, which makes them natural quantum stabilizer codes where their ground states are considered as code subspaces. They are inherently error-correcting, though their multi-qubit syndrome operators are costly to measure. The introduction of the Floquet code \cite{hastings_dynamically_2021} offers an explicit approach to mitigating the complexity of measuring multi-qubit syndrome operators by periodically measuring two-qubit operators, which has been shown to have good error correction properties \cite{gidney_benchmarking_2022}\cite{gidney_fault-tolerant_2021}\cite{paetznick_performance_2023-1}.

One interesting aspect of the Floquet code is its instantaneous phase. At each round of measuring checks, the Floquet state is stabilized by a round-varying stabilizer, the Instantaneous Stabilizer Group (ISG). The original honeycomb code exhibits an instantaneous topological $\mathbf{Z}_2$ phase on a honeycomb lattice, further generalized to any 2D trivalent and 3-colorable planar lattices \cite{vuillot_planar_2021}. In 3D, based on the idea of coupling layers \cite{vijay_isotropic_2017}\cite{ma_fracton_2017}, 3D $X$-cube Floquet codes \cite{zhang_x-cube_2022} and 3D toric code Floquet codes \cite{dua_engineering_2024} have been introduced. Moreover, the latter introduced the rewinding technique to ensure that all instantaneous phases of the Floquet code are topological. On the other hand, the original honeycomb code \cite{hastings_dynamically_2021} exhibits a self-automorphism of instantaneous topological phases. This idea has been further explored through the concept of dynamical automorphism codes \cite{aasen_adiabatic_2022}, where Floquet codes are constructed from adiabatic paths of gapped Hamiltonians. A 3D Floquet color code has been demonstrated based on this idea \cite{davydova_quantum_2024}. These aforementioned codes are always associated with a parent subsystem code, or equivalently can be described by a Kitaev spin liquid model with varying parameters. However, Floquet codes without parent subsystem codes are also possible, as shown in \cite{davydova_floquet_2023}\cite{kesselring_anyon_2024}.

Inspired by the coupling layers construction, we recognized that it inherently possesses the characteristics of coupled spin chains. We present an explicit construction of Floquet codes using coupling spin chains, applicable to a large family of general lattices in any dimension greater than 2, as announced in Section~\ref{sec:3d construction}. It is shown that the previous requirement of trivalent and 3-colorable lattices can be unified into a vertex-2-colorable physical lattice. We provide two different constructions based on the placement of coupling spin chains, which exhibit instantaneous $(n,1)$ toric code topological phases \cite{freedman_double_2016}  and $n$-dimensional $X$-cube fractonic phases, respectively, while the latter agrees with the construction by coupling layers in 3D \cite{zhang_x-cube_2022} on a 3D cubic lattice. Our construction also demonstrates that topological phases can be realized through coupling spin chains, which broadens the applicability of coupling spin chains \cite{halasz_fracton_2017}\cite{williamson_type-ii_2021}\cite{hsieh_fractons_2017}. It naturally leads to the $n$-dimensional $X$-cube model, where we show that, on a hypercubic lattice of size $L$, the leading order of ground state degeneracy for the generalized model is given by $2^{n(n-1)\cdot L^{n-2}}$. The excitations are also composed of lineons and hyper-planons. We show that lineons must form $(n-2)$-dimensional multipoles to unlock extended mobility.

In the 3D construction, unlike in the 2D case, the syndrome operators can no longer survive throughout the measurement routine. They appear at certain rounds, persist for several rounds, and are then removed from the ISG. We need to be cautious about this process regarding error correction: errors are detected by observing changes of the measurement outcomes of the syndrome operators, so the measurement outcomes must be obtained at least twice before being removed from the ISG. If this is achieved, it can correct errors occurring during the rounds when the syndrome operators persist, at low error rate. However, errors outside these rounds need to be detected via other sets of syndromes. Therefore, the synchronization of syndrome operators in the ISG is crucial to ensure error detection for all rounds of the Floquet codes. In our spin chain construction, however, we select a Steady Stabilizer Group (SSG) that persists throughout the Floquet routine and enables a unified decoder. We will show that the ISG forming topological phases is not sufficient for error correction in Floquet codes, but SSG forming classical error-correcting codes is. We present a simple 2-step Floquet Bacon-Shor code \cite{bauer_topological_2024} that fits completely within this framework.

The paper is organized as follows.

In Section~\ref{sec:2d construction}, we first present a clean formulation of the Floquet code in 2D, demonstrating how the 3-colorability of the interactions naturally arises on any 2D lattice. In Section~\ref{subsec:2dconstructions}, we give a detailed explanation of when and how we can get the effective measurement value of syndrome operators so that we can continue the error correction in section~\ref{subsec:2derror}.

We then extend this construction to three dimensions in Section~\ref{sec:3d construction}, where we place one spin chain on each plaquette and show how this generates the 3D Floquet toric code on a cubic lattice. Next, we focus on error correction in Section~\ref{subsec:3d error correction}, which differs from previous works to accommodate general lattices. The error correction relies on a subgroup of the Instantaneous Stabilizer Group (ISG), called the Steady Stabilizer Group (SSG), which persists throughout the measurement routine. In Section~\ref{subsec:IStopological}, we explain that the Floquet code remains error-correctable when the SSG forms a classical error-correcting code at each round. The Floquet Bacon-Shor code is presented as an example that fits this paradigm. This method can be easily generalized to higher-dimensional lattices.

In Section~\ref{subsec:Xcubefc}, we place closed spin chains around vertices, rather than on plaquettes, to construct the $X$-cube Floquet code, which can be generalized to higher-dimensional, locally hypercubic-like lattices. The properties of the generalized $X$-cube model are also discussed.

Finally, in Section~\ref{sec:conclusion}, we conclude the paper and provide a discussion on error correction, automorphisms of Floquet codes, and future directions for research.

Appendix A explicitly shows the instantaneous phases of the Floquet code, which are of particular interest. Appendix B briefly review the Laurent polynomial method and presents a special 3D Kitaev Spin Liquid model on a trivalent lattice that contains a topological phase similar to the 3D toric code. Appendix C analyzes the properties of the generalized $X$-cube code on higher-dimensional hypercubic lattices with periodic boundary conditions.

\section{2D Floquet Code from Spin Chain Construction} \label{sec:2d construction}
\subsection{Spin Chain}
A spin chain is a one-dimensional chain of spins, or qubits, of length $2n$, where nearest-neighbor interactions alternate between $X \otimes X$ and $Y \otimes Y$ terms:

\begin{equation} \label{eqn}
H = -\sum_{k=0}^{n-1} \left(X_{2k} \otimes X_{2k+1} + Y_{2k+1} \otimes Y_{2k+2}\right), \end{equation}
where $X$, $Y$, and $Z$ represent the Pauli matrices, also denoted as $\sigma_x$, $\sigma_y$, and $\sigma_z$:

\begin{equation}
X = \sigma_x = \begin{pmatrix} 0 & 1 \\ 1 & 0 \end{pmatrix},\quad
Y = \sigma_y = \begin{pmatrix} 0 & -i \\ i & 0 \end{pmatrix},\quad
Z = \sigma_z = \begin{pmatrix} 1 & 0 \\ 0 & -1 \end{pmatrix}
\end{equation}
These matrices satisfy the following relation:

\begin{equation} \sigma_i \cdot \sigma_j = \delta_{ij} + i\epsilon_{ijk} \sigma_k, \end{equation}
where $\delta_{ij}$ is the Kronecker delta and $\epsilon_{ijk}$ is the Levi-Civita symbol.

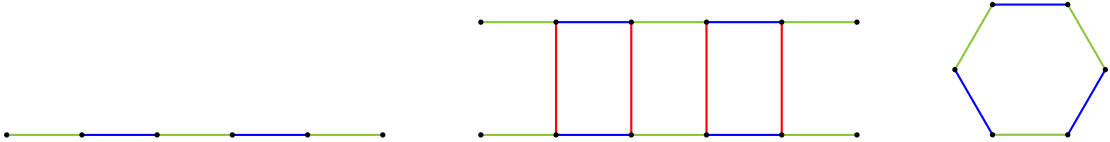
\begin{figure}[htbp]
    \centering
    \begin{tikzpicture}[scale=0.5] 
        \coordinate (A) at (0,0);
        \coordinate (B) at (2,0);
        \coordinate (C) at (4,0);
        \coordinate (D) at (6,0);
        \coordinate (E) at (8,0);
        \coordinate (F) at (10,0);
        
        \draw[thick, LimeGreen] (A) -- (B);
        \draw[thick, blue] (B) -- (C);
        \draw[thick, LimeGreen] (C) -- (D);
        \draw[thick, blue] (D) -- (E);
        \draw[thick, LimeGreen] (E) -- (F);

        \fill[black] (A) circle (2pt);
        \fill[black] (B) circle (2pt);
        \fill[black] (C) circle (2pt);
        \fill[black] (D) circle (2pt);
        \fill[black] (E) circle (2pt);
        \fill[black] (F) circle (2pt);
        
    \end{tikzpicture}
    \hspace{1cm} 
    \begin{tikzpicture}[scale=0.5] 
        \foreach \i in {0,4,8}
        \foreach \j in {0,3}
            \draw[thick, LimeGreen] ({\i},{\j}) -- ({\i+2},{\j});
        \foreach \i in {2,6}
        \foreach \j in {0,3}
            \draw[thick, blue] ({\i},{\j}) -- ({\i+2},{\j});
        
        \foreach \i in {2,4,6,8}
            \draw[thick, red] ({\i},0) -- ({\i},3);
        
        \foreach \i in {0,2,4,6,8,10}
        \foreach \j in {0,3}
            \fill[black] ({\i},{\j}) circle (2pt);
            
    \end{tikzpicture}
    \hspace{1cm} 
    \begin{tikzpicture}[scale=0.5] 
        \coordinate (A) at (0:2);
        \coordinate (B) at (60:2);
        \coordinate (C) at (120:2);
        \coordinate (D) at (180:2);
        \coordinate (E) at (240:2);
        \coordinate (F) at (300:2);
        
        \draw[thick, LimeGreen] (A) -- (B);
        \draw[thick, blue] (B) -- (C);
        \draw[thick, LimeGreen] (C) -- (D);
        \draw[thick, blue] (D) -- (E);
        \draw[thick, LimeGreen] (E) -- (F);
        \draw[thick, blue] (F) -- (A);

        \fill[black] (A) circle (2pt);
        \fill[black] (B) circle (2pt);
        \fill[black] (C) circle (2pt);
        \fill[black] (D) circle (2pt);
        \fill[black] (E) circle (2pt);
        \fill[black] (F) circle (2pt);
    \end{tikzpicture}
    \caption{A demonstration of three different interaction diagrams of spin chains. The left diagram represents an open chain that contains nearest-neighbor $X \otimes X$ and $Y \otimes Y$ interactions, marked by green and blue edges, respectively. The middle diagram shows two spin chains coupled by $Z \otimes Z$ inter-chain check operators, marked by red edges. The right diagram represents a closed hexagonal spin chain.}
    \label{fig:spinchain_threeparts}
\end{figure}

The $Z \otimes Z$ operator is introduced to couple between spin chains, as shown in Figure~\ref{fig:spinchain_threeparts}. Qubits are placed on the black dots. These two-body interactions are called check operators. Specifically, $X \otimes X$ and $Y \otimes Y$ are nearest-neighbor interactions within the same spin chain, referred to as inner-chain check operators, depicted by green and blue edges. On the other hand, $Z \otimes Z$ couples the spin chains and is represented by red edges, named inter-chain operators. Note that all the edges shown in the figure represent interactions, rather than ``real'' lattice edges. We will refer to these as checks, and the figure composed of these checks will be called the interaction diagram throughout this paper.

\subsection{2D Floquet Code}\label{sec:2dfc}
\begin{figure}[htbp]
    \centering
    
    \begin{subfigure}[b]{0.45\textwidth}
        \centering
        \begin{tikzpicture}[scale=0.4]
\foreach \i in {-4, 0, 4}
        \draw[thick, black] ({\i},-7) -- ({\i},6);
\foreach \j in {-4, 0, 4}
        \draw[thick, black] (-6,{\j}) -- (7,{\j});
\foreach \i in {-2, 2}
\foreach \j in {-2, 2}
        \filldraw[fill=red!15, draw=none] ({\i+1},{\j})--({\i},{\j+1})--({\i-1},{\j})--({\i},{\j-1})--cycle;
\draw[thick, LimeGreen] (-2,1) -- (-1,2);
\draw[thick, LimeGreen] (2,1) -- (1,2);
\draw[thick, LimeGreen] (-3,2) -- (-2,3);
\draw[thick, LimeGreen] (3,2) -- (2,3);
\draw[thick, LimeGreen] (-2,-1) -- (-1,-2);
\draw[thick, LimeGreen] (2,-1) -- (1,-2);
\draw[thick, LimeGreen] (-3,-2) -- (-2,-3);
\draw[thick, LimeGreen] (3,-2) -- (2,-3);

\draw[thick, blue] (-2,1) -- (-3,2);
\draw[thick, blue] (2,1) -- (3,2);
\draw[thick, blue] (-1,2) -- (-2,3);
\draw[thick, blue] (1,2) -- (2,3);
\draw[thick, blue] (-2,-1) -- (-3,-2);
\draw[thick, blue] (2,-1) -- (3,-2);
\draw[thick, blue] (-1,-2) -- (-2,-3);
\draw[thick, blue] (1,-2) -- (2,-3);

\draw[thick, blue] (1,-6) -- (2,-5);
\draw[thick, blue] (-1,-6) -- (-2,-5);
\draw[thick, blue] (6,1) -- (5,2);
\draw[thick, blue] (6,-1) -- (5,-2);
\draw[thick, red] (2,3) -- (2,5);
\draw[thick, LimeGreen] (5,2) -- (6,3);

\foreach \j in {-6, -2, 2}
        \draw[thick, red] (-1,{\j}) -- (1,{\j});
\foreach \j in {-2, 2}
        \draw[thick, red] (3,{\j}) -- (5,{\j});
\foreach \i in {-2, 2, 6}
        \draw[thick, red] ({\i},-1) -- ({\i},1);
\foreach \i in {-2, 2}
        \draw[thick, red] ({\i},-5) -- ({\i},-3);

\filldraw[thick, blue] (0,4) circle (6pt);
\filldraw[thick, blue] (0,-4) circle (6pt);
\filldraw[thick, blue] (4,0) circle (6pt);
\filldraw[thick, blue] (-4,0) circle (6pt);
\filldraw[thick, LimeGreen] (0,0) circle (6pt);
\foreach \i in {-4, 4}
\foreach \j in {-4, 4}
        \filldraw[thick, LimeGreen] ({\i},{\j}) circle (6pt);

\end{tikzpicture}
        \caption{}
        \label{subfig:2dsquare}
    \end{subfigure}
    \hfill
    \begin{subfigure}[b]{0.45\textwidth}
        \centering
        \begin{tikzpicture}[scale=0.3]
\draw[thick, blue] (-9,4) -- (-7.5,5);
\draw[thick, blue] (-9,2) -- (-7.5,1);
\draw[thick, blue] (-6,2) -- (-6,4);
\draw[thick, LimeGreen] (-7.5,5) -- (-6,4);
\draw[thick, LimeGreen] (-6,2) -- (-7.5,1);
\draw[thick, LimeGreen] (-9,2) -- (-9,4);
\draw[thick, red] (-7.5,5) -- (-7.5,6.5);
\draw[thick, red] (-6,2) -- (-5,1);
\draw[thick, red] (-9,2) -- (-10.5,1);
\draw[thick, red] (-4,5) -- (-6,4);
\draw[thick, red] (-7.5,-0.5) -- (-7.5,1);
\draw[thick, red] (-10.5,5) -- (-9,4);

\draw[thick, LimeGreen] (-4,5) -- (-4,6.5);
\draw[thick, red] (-6,8) -- (-4,6.5);
\draw[thick, blue] (-2,8) -- (-4,6.5);
\draw[thick, LimeGreen] (-2,8) -- (0,7);
\draw[thick, red] (2,9) -- (0,7);
\draw[thick, blue] (0,3.5) -- (0,7);
\draw[thick, LimeGreen] (0,3.5) -- (-1,3);
\draw[thick, blue] (-4,5) -- (-1,3);
\draw[thick, red] (-2,2) -- (-1,3);
\draw[thick, blue] (-2,2) -- (-5,1);
\draw[thick, LimeGreen] (-2,2) -- (-1.5,1);
\draw[thick, LimeGreen] (-5,1) -- (-4.5,0);
\draw[thick, blue] (-1.5,1) -- (-4.5,0);
\draw[thick, red] (-5,-1) -- (-4.5,0);
\draw[thick, red] (-1.5,1) -- (0,0);
\draw[thick, LimeGreen] (-5,-1) -- (-7.5,-0.5);
\draw[thick, blue] (-10,-2) -- (-7.5,-0.5);
\draw[thick, red] (-10,-2) -- (-12,-1);
\draw[thick, LimeGreen] (-10,-2) -- (-11,-4);
\draw[thick, red] (-13,-5) -- (-11,-4);
\draw[thick, blue] (-3.5,-5.5) -- (-11,-4);
\draw[thick, red] (-3.5,-5.5) -- (-3,-7);
\draw[thick, LimeGreen] (-3.5,-5.5) -- (-2,-4);
\draw[thick, blue] (-5,-1) -- (-2,-4);
\draw[thick, red] (0,-3) -- (-2,-4);
\draw[thick, blue] (0,-3) -- (0,0);
\draw[thick, LimeGreen] (0,-3) -- (2,-4);
\draw[thick, red] (3,-6) -- (2,-4);
\draw[thick, blue] (6,1) -- (2,-4);
\draw[thick, red] (6,1) -- (8,1);
\draw[thick, LimeGreen] (6,1) -- (4,3);
\draw[thick, blue] (1,2) -- (4,3);
\draw[thick, LimeGreen] (1,2) -- (0,0);
\draw[thick, red] (1,2) -- (0,3.5);
\draw[thick, red] (4,3) -- (5,6);

\draw[thick, black] (0,2)--(-4,3)--(-6,0.5)--(-2,-2.5)--cycle;
\draw[thick, black] (0,-6)--(7,-3)--(7,4)--(3,5)--(0,9)--(-4,9)--(-6,6)--(-9,6)--(-11,3)--(-9,0)--(-13,-3)--(-11,-6)--cycle;
\draw[thick, black] (0,2)--(3,5);
\draw[thick, black] (-4,3)--(-6,6);
\draw[thick, black] (-6,0.5)--(-9,0);
\draw[thick, black] (-2,-2.5)--(0,-6);

\filldraw[LimeGreen] (0,2) circle (6pt);
\filldraw[LimeGreen] (0,9) circle (6pt);
\filldraw[LimeGreen] (0,-6) circle (6pt);
\filldraw[LimeGreen] (7,4) circle (6pt);
\filldraw[LimeGreen] (-6,6) circle (6pt);
\filldraw[LimeGreen] (-6,0.5) circle (6pt);
\filldraw[LimeGreen] (-11,3) circle (6pt);
\filldraw[LimeGreen] (-13,-3) circle (6pt);
\filldraw[blue] (7,-3) circle (6pt);
\filldraw[blue] (3,5) circle (6pt);
\filldraw[blue] (-4,9) circle (6pt);
\filldraw[blue] (-4,3) circle (6pt);
\filldraw[blue] (-9,6) circle (6pt);
\filldraw[blue] (-9,0) circle (6pt);
\filldraw[blue] (-2,-2.5) circle (6pt);
\filldraw[blue] (-11,-6) circle (6pt);

        \end{tikzpicture}
        \caption{}
        \label{subfig:2dgeneral}
    \end{subfigure}

    \caption{The two figures above show two physical lattices connected by black edges. The vertices of the lattices are colored green and blue, with no adjacent vertices sharing the same color. In both cases, a closed spin chain is placed on each plaquette, and spin chains are coupled by red edges ($Z \otimes Z$) where they meet at a black edge. The interaction diagram is represented by the colored edges, and one qubit is placed at each vertex of the interaction diagram. Figure~\ref{subfig:2dsquare} depicts a square lattice, while Figure~\ref{subfig:2dgeneral} shows a general planar lattice. It is clear that as long as the vertices connected by black edges can be 2-colored, the interaction diagram forms a trivalent, 3-colorable lattice, which has been shown to support a planar Floquet code \cite{vuillot_planar_2021}.}
    \label{fig:2dinteractiondiagram}
\end{figure}
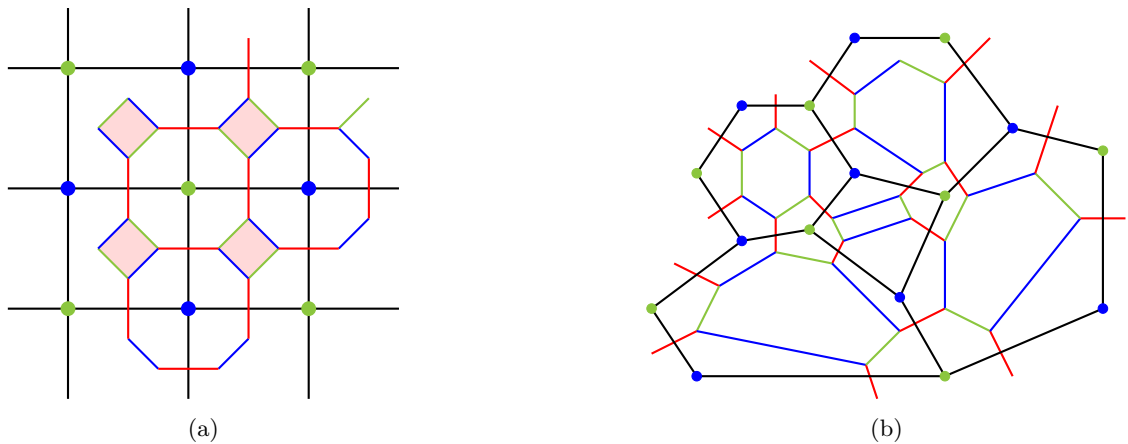

We begin by restating the 2D Floquet code for further generalization. Throughout this paper, we will only consider lattices with periodic boundary conditions unless specifically mentioned otherwise. Consider a 2D physical lattice $\Gamma$, represented by the lattice connected by black edges in Figure~\ref{fig:2dinteractiondiagram}. Let $E$, $V$, and $P$ represent the sets of edges, vertices, and plaquettes, respectively. For each plaquette $p$ of the physical lattice, we place a closed spin chain that turns at the boundary edges, with one qubit located at each turning point. We slightly deform the turning points away from the edges to emphasize the independence of the qubits from different spin chains, though it is important to note that these qubits are still placed on the physical edges.

As a simple example, we take the 2D square lattice shown in Figure~\ref{subfig:2dsquare}, where black edges represent the physical lattice. One closed spin chain is placed on each plaquette. Since each black edge borders two plaquettes, two spin chains will coincide at the edge from different directions. In this case, we place a $Z \otimes Z$ operator, marked by red edges, to couple the two qubits that coincide at the same edge.

The edge-colored lattice represents the interaction diagram, denoted as $\Gamma'$. This lattice is automatically trivalent in 2D, as each edge borders only two plaquettes. Interestingly, when the original lattice $\Gamma$ is 2-colorable at all vertices, $\Gamma'$ becomes a trivalent and 3-colorable lattice, which has been shown to support a planar Floquet code in \cite{vuillot_planar_2021}.

In the figure, we match the edge colors in $\Gamma'$ to the vertex colors of $\Gamma$ as follows: 
- All edges of $\Gamma'$ crossing a black edge are colored red.
- All edges of $\Gamma'$ contained within plaquettes of $\Gamma$ are colored according to the nearest vertices of $\Gamma$. The plaquettes of $\Gamma'$ are then colored as:
  - Green if bordered only by green and red checks,
  - Blue if bordered only by blue and red checks,
  - Red if bordered only by blue and green checks.

We break the 3-color symmetry here, as this symmetry naturally breaks in higher dimensions. The interaction diagram obtained from the physical square lattice in Figure~\ref{subfig:2dsquare} explicitly represents the 2D Floquet code on a 4.8.8 lattice. In the original lattice $\Gamma$, the red-colored plaquettes correspond to the original plaquettes, while the green and blue plaquettes correspond to the two-colored vertices of $\Gamma$. 

Now we can conclude that on a 2D lattice where the vertices are 2-colorable, we can place closed spin chains over the plaquettes, which will support a 2D Floquet code. Specifically, immediately after the measurement of the red checks, or at the strong coupling limit, the instantaneous phase corresponds explicitly to the toric code on $\Gamma$, where qubits are placed on the edges of $\Gamma$. 

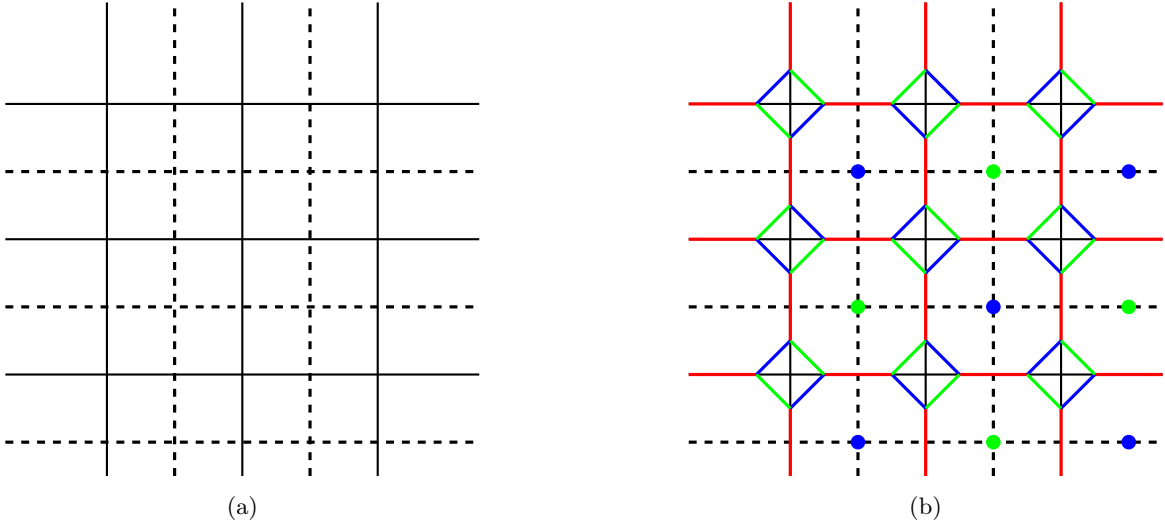
\begin{figure}[htbp]
    \centering
    \begin{subfigure}[t]{0.45\textwidth}
        \centering
        \begin{tikzpicture}[scale=0.45]
            \foreach \i in {-4,0,4}
                \draw[thick, black] ({\i},-7)--({\i},7);
            \foreach \j in {-4,0,4}
                \draw[thick, black] (-7,{\j})--(7,{\j});
            \foreach \i in {-2,2}
                \draw[very thick, dashed, black] ({\i},-7)--({\i},7);
            \foreach \j in {-6,-2,2}
                \draw[very thick, dashed, black] (-7,{\j})--(7,{\j});
        \end{tikzpicture}
        \caption{}
        \label{subfig:2dduallattice}
    \end{subfigure}
    \hfill
    \begin{subfigure}[t]{0.45\textwidth}
        \centering
        \begin{tikzpicture}[scale=0.45]
            \foreach \i in {-4,0,4}
                \draw[thick, black] ({\i},-7)--({\i},7);
            \foreach \j in {-4,0,4}
                \draw[thick, black] (-7,{\j})--(7,{\j});
            \foreach \i in {-2,2}
                \draw[very thick, dashed, black] ({\i},-7)--({\i},7);
            \foreach \j in {-6,-2,2}
                \draw[very thick, dashed, black] (-7,{\j})--(7,{\j});
            \foreach \i in {-7,-3,1,5}
            \foreach \j in {-4,0,4}
                \draw[very thick, red] ({\i},{\j})--({\i+2},{\j});
            \foreach \j in {-7,-3,1,5}
            \foreach \i in {-4,0,4}
                \draw[very thick, red] ({\i},{\j})--({\i},{\j+2});
            
            \draw[very thick, blue] (0,-1)--(1,0);
            \draw[very thick, blue] (0,1)--(-1,0);
            \foreach \j in {-5,3}
            \foreach \i in {-4,4}
                \draw[very thick, blue] ({\i},{\j})--({\i+1},{\j+1});
            \foreach \j in {-3,5}
            \foreach \i in {-4,4}
                \draw[very thick, blue] ({\i},{\j})--({\i-1},{\j-1});
            \foreach \j in {-3,5}
                \draw[very thick, blue] (0,{\j})--(1,{\j-1});
            \foreach \j in {-5,3}
                \draw[very thick, blue] (0,{\j})--(-1,{\j+1});
            \foreach \i in {-4,4}
                \draw[very thick, blue] ({\i},{1})--({\i+1},{0});
            \foreach \i in {-4,4}
                \draw[very thick, blue] ({\i},{-1})--({\i-1},{0});
            
            \draw[very thick, green] (0,-1)--(-1,0);
            \draw[very thick, green] (0,1)--(1,0);
            \foreach \j in {-5,3}
            \foreach \i in {-4,4}
                \draw[very thick, green] ({\i},{\j})--({\i-1},{\j+1});
            \foreach \j in {-3,5}
            \foreach \i in {-4,4}
                \draw[very thick, green] ({\i},{\j})--({\i+1},{\j-1});
            \foreach \j in {-3,5}
                \draw[very thick, green] (0,{\j})--(-1,{\j-1});
            \foreach \j in {-5,3}
                \draw[very thick, green] (0,{\j})--(1,{\j+1});
            \foreach \i in {-4,4}
                \draw[very thick, green] ({\i},{-1})--({\i+1},{0});
            \foreach \i in {-4,4}
                \draw[very thick, green] ({\i},{1})--({\i-1},{0});
            \filldraw[blue] (2,-2) circle (0.2);
            \foreach \i in {-2,6}
            \foreach \j in {-6,2}
                \filldraw[blue] ({\i},{\j}) circle (0.2);
            \filldraw[green] (2,-6) circle (0.2);
            \filldraw[green] (2,2) circle (0.2);
            \filldraw[green] (-2,-2) circle (0.2);
            \filldraw[green] (6,-2) circle (0.2);
        \end{tikzpicture}
        \caption{}\label{subfig:fcondual}
    \end{subfigure}
    \caption{Figure~\ref{subfig:2dduallattice} shows the dual lattice $\Gamma'$ of a square lattice $\Gamma$. $\Gamma$ is connected by thin black edges, and $\Gamma'$ by dashed edges. $\Gamma$ and $\Gamma'$ are dual to each other. Figure~\ref{subfig:fcondual} illustrates the case where the closed spin chains are placed on the plaquettes of $\Gamma'$, or equivalently around the vertices of $\Gamma$. Note that in this case, the inner-chain checks are colored based on the 2-coloring of the plaquettes of $\Gamma$.}
    \label{fig:2ddual}
\end{figure}

Alternatively, we can place closed spin chains on the faces of the dual lattice $\bar{\Gamma}$ of $\Gamma$, which can also be viewed as placing them on the vertices of $\Gamma$, as shown in Figure~\ref{fig:2ddual}. The 3-coloring of the new interaction diagram is determined by the 2-coloring of the vertices of $\bar{\Gamma}$, or equivalently, the 2-coloring of the faces of $\Gamma$. 

It is clear that these two different placements of closed spin chains are equivalent due to the duality of the 2D toric code, and they yield equivalent 2D Floquet codes. However, as we will see, in higher dimensions, these two placements lead to different outcomes.

\subsection{Referred Syndrome Operators}\label{subsec:2dconstructions}
The essence of the Floquet code is to avoid directly measuring the syndrome operators, although we still need the measurement results of syndromes for error detection and correction. Here, we provide a detailed explanation of how and when the measurement outcomes are inferred from the measurement results of check operators.

With the coupled spin chain construction, the Floquet routine proceeds as follows: at step $3r$, we measure the red checks; at step $3r + 1$, we measure the blue checks; and at step $3r + 2$, we measure the green checks, as shown in Table~\ref{tab:2d_floquet_routine}. Suppose the green check operators are labeled as $O_g$, and the projectors onto the eigenstate with eigenvalue $+1$ are labeled as $P_g$. Similarly, blue and red checks are associated with $O_b$, $O_r$, and $P_b$, $P_r$, respectively. Starting with an initial state $|\phi\rangle$, the state after round 2 can be written as:

\begin{equation}
    |\phi_{2}\rangle =  P_g \cdot P_b \cdot  P_r \cdots |\phi\rangle 
\end{equation}\label{eqn:2d_floquet_state}

We abbreviate the notation so that each $P_i$ in the above equation, where $i \in \{r, b, g\}$, represents the product of all projectors of the same color. Here, we assume that all measurement results are $+1$, although other outcomes can be treated equivalently.

To better understand how the measurement outcome of a red plaquette operator is inferred from the measurement outcomes of the checks, consider the example shown in Figure~\ref{fig:2dinteractiondiagram}. Take a red plaquette that is bordered by two green checks, two blue checks, and four red checks attached to its boundary, each labeled by a number. We can then rewrite the state $|\phi_2\rangle$ at round 2 as:

\begin{equation}
    |\phi_{2}\rangle = P_{g1} P_{g2} \cdot P_{b1} P_{b2} \cdot P_{r1} P_{r2} P_{r3} P_{r4} |\phi\rangle
\end{equation}

Note that we neglect all other projectors that have no common support with this red plaquette operator in the above equation. The plaquette operators are denoted by $W_i$, where each $W_i$ is the product of the check operators on the boundary of a plaquette of color $i$, with $i \in \{r, b, g\}$. For this example, a red plaquette can be written explicitly as $W_r = O_{g1} O_{b1} O_{g2} O_{b2}$. When applied to the state $|\phi_{3r+2}\rangle$, we have:

\begin{align}
    & O_{g1} O_{b1} O_{g2} O_{b2} P_{g1} P_{g2} \cdot P_{b1} P_{b2} \cdot P_{r1} P_{r2} P_{r3} P_{r4} |\phi\rangle \notag \\
    &= - O_{g1} O_{g2} O_{b1} O_{b2} P_{g1} P_{g2} \cdot P_{b1} P_{b2} \cdot P_{r1} P_{r2} P_{r3} P_{r4} |\phi\rangle \notag \\
    &= - O_{g1} O_{g2} P_{g1} P_{g2} \cdot O_{b1} O_{b2} P_{b1} P_{b2} \cdot P_{r1} P_{r2} P_{r3} P_{r4} |\phi\rangle \notag \\
    &= - P_{g1} P_{g2} \cdot P_{b1} P_{b2} \cdot P_{r1} P_{r2} P_{r3} P_{r4} |\phi\rangle
\end{align}\label{eqn:referred_syndrome}

The equation above holds because, for any operator $O$, we have $O^2 = \text{Id}$, and the corresponding projector onto the $+1$ eigenstate is $P = \frac{1 + O}{2}$, thus $OP = P$. Additionally, $O_{b1} O_{b2}$ can pass through $P_g$ from the second to the third line, since the product of the blue checks around a red plaquette commutes with the green checks.

The entire Floquet state is the eigenstate of $W_r$ with eigenvalue $-1$, and its value is determined, or we say ``inferred'', from the measurements of check operators, immediately after the measurement of the green checks at step $3r + 2$, where $r \in \mathbf{Z}$. This process works similarly for other plaquette operators. The plaquette operators form the Steady Stabilizer Group (SSG), a subgroup of the Instantaneous Stabilizer Group (ISG). We simply have $\text{ISG}_r = \text{SSG} \cup \{r\text{-checks}\}$. Elements in SSG will have fixed values if no error occurs and can be effectively ``measured'' in the Floquet code, enabling error detection by reading their updated measurement outcomes.

\begin{table}[htbp]
    \centering
    \begin{tabular}{|c|c|c|c|c|c|}
        \textbf{Steps} & \textbf{3r} & \textbf{3r+1} & \textbf{3r+2} & \textbf{3r+3} & \textbf{\dots} \\
        Measure checks & Red & Blue & Green & Red & \textbf{\dots} \\
        Updated stabilizers & Blue & Green & Red & Blue & \textbf{\dots} \\
    \end{tabular}
    \caption{Measurement routine of the 2D Floquet code. At each step, one type of plaquette operator is referred, allowing the measurement result of the syndrome operator to be updated without directly measuring it.}
    \label{tab:2d_floquet_routine}
\end{table}


\subsection{Error Correction}\label{subsec:2derror}
Note that elements in the Steady Stabilizer Group (SSG) survive throughout the measurement routine and are updated periodically. As a result, they can be used as syndrome operators to detect the occurrence of errors. Here, we briefly outline the error correction method, which is largely similar to the approach presented in \cite{hastings_dynamically_2021}.

The interaction diagram has a trivalent nature, meaning any error must be surrounded by exactly three plaquettes with distinct actions on the qubit. Thus, any Pauli error will anticommute with two of the plaquette operators.

\begin{figure}[htbp]
    \centering
    \begin{tikzpicture}[scale=0.5]
        \foreach \i in {-4,0,4}
                \draw[very thick, black] ({\i},-7)--({\i},6);
        \foreach \j in {-4,0,4}
                \draw[very thick, black] (-7,{\j})--(6,{\j});
        \foreach \i in {-6,-2,2,6}
        \foreach \j in {-5,-1,3}
                \draw[very thick, red] ({\i},{\j})--({\i},{\j+2});
        \foreach \j in {-6,-2,2,6}
        \foreach \i in {-5,-1,3}
                \draw[very thick, red] ({\i},{\j})--({\i+2},{\j});
        \draw[very thick, blue] (1,-2)--(2,-1);
        \draw[very thick, blue] (1,2)--(2,1);
        \draw[very thick, blue] (-1,-2)--(-2,-1);
        \draw[very thick, blue] (-1,2)--(-2,1);
        \foreach \i in {-6,2}
        \foreach \j in {-5,3}
                \draw[very thick, blue] ({\i},{\j})--({\i+1},{\j-1});
        \foreach \i in {-6,2}
        \foreach \j in {-3,5}
                \draw[very thick, blue] ({\i},{\j})--({\i+1},{\j+1});
        \foreach \i in {-2,6}
        \foreach \j in {-5,3}
                \draw[very thick, blue] ({\i},{\j})--({\i-1},{\j-1});
        \foreach \i in {-2,6}
        \foreach \j in {-3,5}
                \draw[very thick, blue] ({\i},{\j})--({\i-1},{\j+1});
        \foreach \i in {-6,2}
                \draw[very thick, green] ({\i},-1)--({\i+1},-2);
        \foreach \i in {-6,2}
                \draw[very thick, green] ({\i},1)--({\i+1},2);
        \foreach \i in {-2,6}
                \draw[very thick, green] ({\i},-1)--({\i-1},-2);
        \foreach \i in {-2,6}
                \draw[very thick, green] ({\i},1)--({\i-1},2);
        \foreach \j in {-6,2}
                \draw[very thick, green] (-1,{\j})--(-2,{\j+1});
        \foreach \j in {-6,2}
                \draw[very thick, green] (1,{\j})--(2,{\j+1});
        \foreach \j in {-2,6}
                \draw[very thick, green] (-1,{\j})--(-2,{\j-1});
        \foreach \j in {-2,6}
                \draw[very thick, green] (1,{\j})--(2,{\j-1});
        
        \filldraw[blue] (0,0) circle (0.2); 
        \foreach \i in {-4,4}
        \foreach \j in {-4,4}
                \filldraw[blue] ({\i},{\j}) circle (0.2);
        \foreach \i in {-4,4}
                \filldraw[green] ({\i},0) circle (0.2);
        \foreach \j in {-4,4}
                \filldraw[green] (0,{\j}) circle (0.2);
        
        \draw[very thick, green] (-1,2) circle (0.4); 

        \draw[ultra thick, green] (-1,2)--(-2,3); 

        \node at (-1.6, 1.6) {\textbf{r}};

        \node at (1, 3.25) {\textbf{b}};
        
        \filldraw[red] (-2,2) circle (0.2); 
        \filldraw[red] (2,2) circle (0.2);  
        \filldraw[red] (-2,-2) circle (0.2); 
        \filldraw[red] (2,-2) circle (0.2);  
    \end{tikzpicture}
    \caption{This figure shows the colored interaction diagram of the 2D Floquet code over a physical black square lattice. Each colored edge represents a check operator. Each plaquette is colored based on the color of its center dot and is associated with a plaquette operator. When a green error, marked by a green circle, occurs, the two plaquette operators on the plaquettes marked by $r$ and $b$ will be flipped. Conversely, when the values of these two plaquette operators are flipped, we know that a green error occurred at the thickened green edge. We can simply apply a green operator to either end of the green edge, which will either correct the error or create a green check error that will disappear automatically.}
    \label{fig:2derror}
\end{figure}
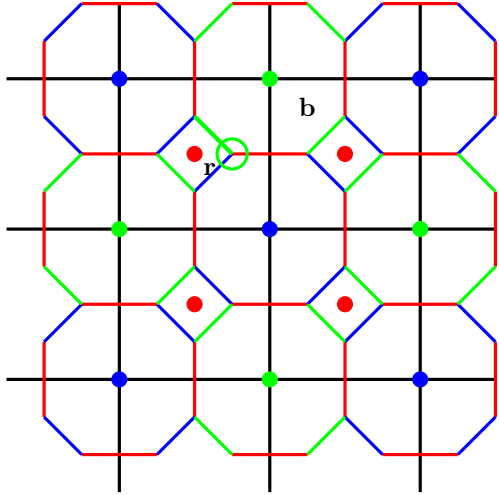

We refer to Pauli $X$, $Y$, and $Z$ errors as green, blue, and red errors, respectively. Any single Pauli error can be detected by observing that two measurement results have flipped. For example, a green error, as shown in Figure~\ref{fig:2derror}, will flip the nearest green and red plaquette operators. Conversely, if the measurement of one red and one green plaquette operator flips, we can deduce that a green error occurred at the shared edge, the thickened edge in the figure, which must be a green edge under our assignment.

It is not possible to distinguish between green errors at either end of a green check, since $X \otimes I = (X \otimes X) \cdot (I \otimes X)$, and plaquette operators always commute with all check operators. However, this does not affect the correctability of the error. We can simply apply a green operator to either end of the green edge. For example, applying $X \otimes I$ to the first vertex will either cancel the error if it was indeed $X \otimes I$, or result in $X \otimes X$ acting on the Floquet state if the error was $I \otimes X$. This $X \otimes X$ is a check error, which will disappear automatically, as pointed out in the original paper, and can also be verified in Equation~\ref{eqn:2d_floquet_state}, since the check operator will contribute only an overall constant after it is measured.

The above argument applies similarly to any Pauli error on any vertex where surrounding plaquette operators are recorded. Thus, the entire Floquet code is error-correctable under a low error rate.

\section{3D Error-Correctable Floquet Code} \label{sec:3d construction}
\subsection{3D Floquet Toric Code}\label{subsec:3d toric code fc}
In two dimensions, there is only one topological phase that can be realized by a frustration-free Pauli Hamiltonian: the 2D toric code. However, in higher dimensions, there are clearly more topological phases. Therefore, we can expect different Floquet codes with distinct instantaneous topological phases arising from the coupling spin chain construction.

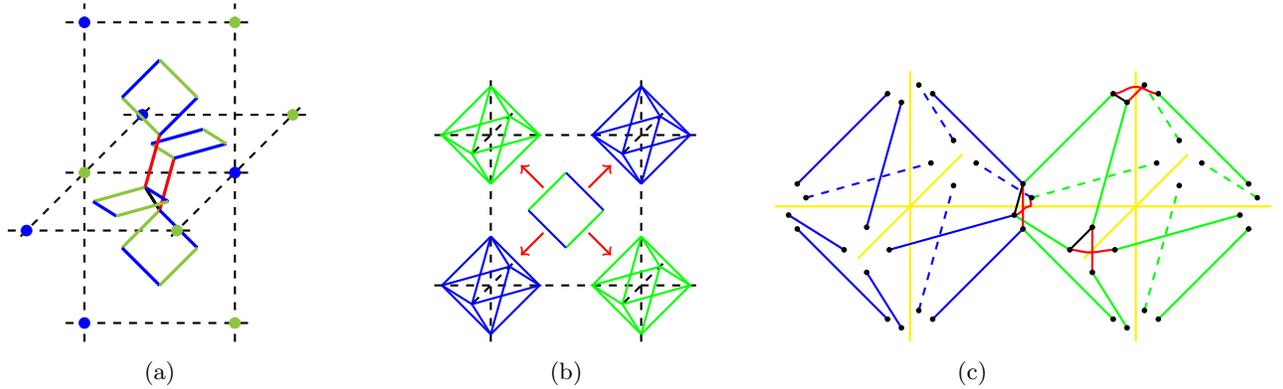
\begin{figure}[htbp]
    \centering
    
    \begin{subfigure}[b]{0.32\textwidth}
        \centering
        \begin{tikzpicture}[scale=0.5]
            \foreach \j in {-4, 0, 4}
                    \draw[thick, black, dashed] (-2.5,{\j},0) -- (2.5,{\j},0);
            \foreach \k in {-4, 4}
                    \draw[thick, black, dashed] (-2.5,0,{\k}) -- (2.5,0,{\k});
            \foreach \i in {-2,  2}
                    \draw[thick, black, dashed] ({\i},0,-4.5) -- ({\i},0,4.5);
            \foreach \i in {-2,  2}
                    \draw[thick, black, dashed] ({\i},-4.5,0) -- ({\i},4.5,0);

            \filldraw[blue] (-2,0,-4) circle (4pt);

            \draw[very thick, black] (0,-1,0) -- (0,0,1);
            \draw[very thick, red] (0,-1,0) -- (0,0,-1);
            \foreach \k in {-1,  1}
                    \draw[very thick, LimeGreen] (-1,0,{\k*2}) -- (0,0,{\k*1});
            \foreach \k in {-1,  1}
                    \draw[very thick, blue] (-1,0,{\k*2}) -- (0,0,{\k*3});
            \foreach \k in {-1,  1}
                    \draw[very thick, blue] (1,0,{\k*2}) -- (0,0,{\k*1});
            \foreach \k in {-1,  1}
                    \draw[very thick, LimeGreen] (1,0,{\k*2}) -- (0,0,{\k*3});

            \foreach \j in {-1,  1}
                    \draw[very thick, LimeGreen] (-1,{\j*2},0) -- (0,{\j*1},0);
            \foreach \j in {-1,  1}
                    \draw[very thick, blue] (-1,{\j*2},0) -- (0,{\j*3},0);
            \foreach \j in {-1,  1}
                    \draw[very thick, blue] (1,{\j*2},0) -- (0,{\j*1},0);
            \foreach \j in {-1,  1}
                    \draw[very thick, LimeGreen] (1,{\j*2},0) -- (0,{\j*3},0);
            \draw[very thick, red] (0,1,0) -- (0,0,1);

            \filldraw[blue] (-2,0,4) circle (4pt);
            \filldraw[blue] (-2,4,0) circle (4pt);
            \filldraw[blue] (-2,-4,0) circle (4pt);
            \filldraw[blue] (2,0,0) circle (4pt);

            \filldraw[LimeGreen] (-2,0,0) circle (4pt);
            \filldraw[LimeGreen] (2,4,0) circle (4pt);
            \filldraw[LimeGreen] (2,-4,0) circle (4pt);
            \filldraw[LimeGreen] (2,0,-4) circle (4pt);
            \filldraw[LimeGreen] (2,0,4) circle (4pt);
        \end{tikzpicture}
        \caption{}
        \label{subfig:3dtoriccube}
    \end{subfigure}
    \begin{subfigure}[b]{0.32\textwidth}
        \centering
        \begin{tikzpicture}[scale=0.5]
            \foreach \i in {-2, 2}
            \foreach \j in {-2, 2}
                    \draw[thick, dashed, black] ({\i},{\j},-1.5) -- ({\i},{\j},1.5);
            \foreach \i in {-2, 2}
                    \draw[thick, dashed, black] ({\i},-3.5,0) -- ({\i},3.5,0);
            \foreach \j in {-2, 2}
                    \draw[thick, dashed, black] (-3.5,{\j},0) -- (3.5,{\j},0);

            \foreach \i in {-1.3, 1.3}
                    \draw[thick, blue] (2,3.3,0) -- ({2+\i},2,0) -- (2,0.7,0);
            \foreach \k in {-1.3, 1.3}
                    \draw[thick, blue] (2,3.3,0) -- (2,2,{\k}) -- (2,0.7,0);
            \draw[thick, blue] (2+1.3,2,0) -- (2,2,0+1.3) -- (2-1.3,2,0) -- (2,2,0-1.3) -- cycle;

            \foreach \i in {-1.3, 1.3}
                    \draw[thick, blue] (-2,-3.3,0) -- ({-2+\i},-2,0) -- (-2,-0.7,0);
            \foreach \k in {-1.3, 1.3}
                    \draw[thick, blue] (-2,-3.3,0) -- (-2,-2,{\k}) -- (-2,-0.7,0);
            \draw[thick, blue] (-2+1.3,-2,0) -- (-2,-2,0+1.3) -- (-2-1.3,-2,0) -- (-2,-2,0-1.3) -- cycle;

            \foreach \i in {-1.3, 1.3}
                    \draw[thick, green] (2,-3.3,0) -- ({2+\i},-2,0) -- (2,-0.7,0);
            \foreach \k in {-1.3, 1.3}
                    \draw[thick, green] (2,-3.3,0) -- (2,-2,{\k}) -- (2,-0.7,0);
            \draw[thick, green] (2+1.3,-2,0) -- (2,-2,0+1.3) -- (2-1.3,-2,0) -- (2,-2,0-1.3) -- cycle;

            \foreach \i in {-1.3, 1.3}
                    \draw[thick, green] (-2,3.3,0) -- ({-2+\i},2,0) -- (-2,0.7,0);
            \foreach \k in {-1.3, 1.3}
                    \draw[thick, green] (-2,3.3,0) -- (-2,2,{\k}) -- (-2,0.7,0);
            \draw[thick, green] (-2+1.3,2,0) -- (-2,2,0+1.3) -- (-2-1.3,2,0) -- (-2,2,0-1.3) -- cycle;

            \draw[thick, green] (-1,0,0) -- (0,1,0);
            \draw[thick, green] (1,0,0) -- (0,-1,0);
            \draw[thick, blue] (-1,0,0) -- (0,-1,0);
            \draw[thick, blue] (0,1,0) -- (1,0,0);
            \draw[->, thick, red] (0.6,0.6,0) -- (1.2,1.2,0);
            \draw[->, thick, red] (-0.6,-0.6,0) -- (-1.2,-1.2,0);
            \draw[->, thick, red] (-0.6,0.6,0) -- (-1.2,1.2,0);
            \draw[->, thick, red] (0.6,-0.6,0) -- (1.2,-1.2,0);
        \end{tikzpicture}
        \caption{}
        \label{subfig:realign}
    \end{subfigure}
    \begin{subfigure}[b]{0.32\textwidth}
        \centering
        \begin{tikzpicture}[scale=0.3]
            \draw[thick, yellow] (-11,0,0)--(11,0,0);
            \foreach \i in {-5, 5}
                    \draw[thick, yellow] ({\i},-6,0)--({\i},6,0);
            \foreach \i in {-5, 5}
                    \draw[thick, yellow] ({\i},0,-6)--({\i},0,6);

            \draw[thick, green] (0,1,0)--(4,5,0);
            \draw[thick, green] (0,-1,0)--(4,-5,0);
            \draw[thick, green] (0,0,1)--(4,0,5);
            \draw[thick, dashed, green] (0,0,-1)--(4,0,-5);
            \draw[thick, green] (10,1,0)--(6,5,0);
            \draw[thick, green] (10,-1,0)--(6,-5,0);
            \draw[thick, green] (10,0,1)--(6,0,5);
            \draw[thick, dashed, green] (10,0,-1)--(6,0,-5);
            \draw[thick, green] (5,5,1)--(5,1,5);
            \draw[thick, green] (5,-1,5)--(5,-5,1);
            \draw[thick, dashed, green] (5,-5,-1)--(5,-1,-5);
            \draw[thick, dashed, green] (5,1,-5)--(5,5,-1);

            \draw[thick, blue] (0,1,0)--(-4,5,0);
            \draw[thick, blue] (0,-1,0)--(-4,-5,0);
            \draw[thick, blue] (0,0,1)--(-4,0,5);
            \draw[thick, dashed, blue] (0,0,-1)--(-4,0,-5);
            \draw[thick, blue] (-10,1,0)--(-6,5,0);
            \draw[thick, blue] (-10,-1,0)--(-6,-5,0);
            \draw[thick, blue] (-10,0,1)--(-6,0,5);
            \draw[thick, dashed, blue] (-10,0,-1)--(-6,0,-5);
            \draw[thick, blue] (-5,5,1)--(-5,1,5);
            \draw[thick, blue] (-5,-1,5)--(-5,-5,1);
            \draw[thick, dashed, blue] (-5,-5,-1)--(-5,-1,-5);
            \draw[thick, dashed, blue] (-5,1,-5)--(-5,5,-1);            
            \draw[thick, red] (5,5,-1)--(5,5,1);
            \draw[thick, smooth, red] (4,5,0) to [out=0, in=180] (5,5.3,0) to [out=0, in=180] (6,5,0);

            \draw[thick, red] (0,1,0)--(0,-1,0);
            \draw[thick, smooth, red] (0,0,1) to [out=0, in=180] (0.3,0,0) to [out=0, in=180] (0,0,-1);

            \draw[thick, red] (5,1,5)--(5,-1,5);
            \draw[thick, smooth, red] (4,0,5) to [out=0, in=180] (5,0,5.3) to [out=0, in=180] (6,0,5);

            \draw [thick, black] (0,1,0)--(0,0,1);
            \draw [thick, black] (4,0,5)--(5,1,5);
            \draw [thick, black] (4,5,0)--(5,5,1);

            \foreach \i in {-10, 0, 10}
            \foreach \k in {1, -1}
                    \filldraw[black] ({\i},0,{\k}) circle (0.1);
            \foreach \i in {-10, 0, 10}
            \foreach \j in {1, -1}
                    \filldraw[black] ({\i},{\j},0) circle (0.1);

            \foreach \s in {-1, 1}
            \foreach \i in {4, 6}
            \foreach \j in {-5, 5}
                    \filldraw[black] ({\s*\i},{\j},0) circle (0.1);
            \foreach \s in {-1, 1}
            \foreach \j in {-5, 5}
            \foreach \k in {-1, 1}
                    \filldraw[black] ({\s*5},{\j},{\k}) circle (0.1);

            \foreach \s in {-1, 1}
            \foreach \i in {4, 6}
            \foreach \k in {-5, 5}
                    \filldraw[black] ({\s*\i},0,{\k}) circle (0.1);
            \foreach \s in {-1, 1}
            \foreach \k in {-5, 5}
            \foreach \j in {-1, 1}
                    \filldraw[black] ({\s*5},{\j},{\k}) circle (0.1);
        \end{tikzpicture}
        \caption{}
        \label{subfig:brokenoctahedron}
        \end{subfigure}

   \caption{The above diagrams show the 3D Floquet toric code. Figure~\ref{subfig:3dtoriccube} depicts a physical cubic lattice $\Gamma$, represented by dashed black edges. A closed qubit chain is placed on each plaquette of $\Gamma$. Note that four qubits are located on each edge of $\Gamma$, though we separate them a little apart for clarity. We introduce three $Z \otimes Z$ check operators to couple them: two colored red and one black, with the two red checks disjoint from each other. In Figure~\ref{subfig:realign}, the inner-chain checks are moved closer to the vertices, forming octahedra centered on the vertices of $\Gamma$. The inner-chain checks are colored according to the color of the vertex at the center of the octahedron they belong to. In the 3D Floquet code, the product of check operators (excluding the black checks) bordering each octahedron forms a stabilizer in the Steady Stabilizer Group. Figure~\ref{subfig:brokenoctahedron} shows an explicit arrangement of the colored checks. In this case, the yellow edges represent the physical edges. For clarity, only three groups of red and black checks are shown, but in reality, these checks are placed on any four qubits adjacent to a physical edge.}
\end{figure}

We take a 3D cubic physical lattice $\Gamma$ as an example, as shown in Figure~\ref{subfig:3dtoriccube}, where the physical cubic lattice is represented by dashed lines. This is a direct generalization of the 2D case shown in Figure~\ref{fig:2dinteractiondiagram}. On each plaquette of the lattice, we place a closed spin chain, and we introduce $Z \otimes Z$ interactions between spin chains where they meet at the same physical edge. As before, this leads to an interaction diagram $\Gamma'$. For convenience in coloring, we slightly deform the figure by aligning the inner-chain links to their nearest physical vertices.

We 2-color the vertices of $\Gamma$, and the edges of $\Gamma'$ (the inner-chain checks) are colored based on their nearest physical vertex. The inter-chain checks are further divided into two parts. One part, colored red, consists of a set of check operators whose product covers all qubits on one edge of $\Gamma$. For example, in the 3D cubic lattice, assigning a closed spin chain to each plaquette results in four qubits surrounding one edge of $\Gamma$. To handle this, we select two non-overlapping inter-chain checks to be red, while the remaining one is marked black, as shown in Figure~\ref{subfig:3dtoriccube}. Note that in 2D, this black inter-chain check naturally disappears because the entire lattice is trivalent.

\begin{table}[htbp]
    \centering
    \begin{tabular}{c c p{5cm}}
        \toprule
        Round & Check measured & ISG \\ \midrule
        6r & Red + Black & SSG + red checks + black checks \\ \midrule
        6r+1 & Green & SSG + triangular-green + green checks \\ \midrule
        6r+2 & Blue & SSG + triangular-green + blue checks \\ \midrule
        6r+3 & Red & SSG + NSSG\_3 + red checks \\ \midrule
        6r+4 & Green & SSG + NSSG\_4 + blue checks \\ \midrule
        6r+5 & Blue & SSG + NSSG\_5 + green checks \\ \bottomrule
    \end{tabular}
    \caption{This table shows the Instantaneous Stabilizer Group (ISG) at each round of the measurement routine, containing the Steady Stabilizer Group (SSG), possible Non-Steady Stabilizer Groups (NSSG), and the check operators measured in the current round. The triangular operators in the table are depicted in Figure~\ref{fig:triangularoperator}. The actual NSSG\_{\{3,4,5\}} is not shown explicitly, as it varies depending on the choice of red checks.}\label{tab:3dISG}
\end{table}

\begin{figure}[htbp]
    \centering
    \begin{subfigure}[t]{0.48\textwidth}
        \centering
        \begin{tikzpicture}[scale=0.37]
            \draw[thick, red] (4,5,0)--(5,5,1);
            \draw[thick, red] (0,1,0)--(0,0,1);
            \draw[thick, red] (4,0,5)--(5,1,5);
            \draw[red] (4,5,0) circle(0.3);
            \draw[red] (5,5,1) circle(0.3);
            \draw[red] (0,1,0) circle(0.3);
            \draw[red] (0,0,1) circle(0.3);
            \draw[red] (4,0,5) circle(0.3);
            \draw[red] (5,1,5) circle(0.3);

            \draw[thick, yellow] (-11,0,0)--(11,0,0);
            \foreach \i in {-5, 5}
                \draw[thick, yellow] ({\i},-6,0)--({\i},6,0);
            \foreach \i in {-5, 5}
                \draw[thick, yellow] ({\i},0,-6)--({\i},0,6);

            \draw[thick, green] (0,1,0)--(4,5,0);
            \draw[thick, green] (0,-1,0)--(4,-5,0);
            \draw[thick, green] (0,0,1)--(4,0,5);
            \draw[thick, dashed, green] (0,0,-1)--(4,0,-5);
            \draw[thick, green] (10,1,0)--(6,5,0);
            \draw[thick, green] (10,-1,0)--(6,-5,0);
            \draw[thick, green] (10,0,1)--(6,0,5);
            \draw[thick, dashed, green] (10,0,-1)--(6,0,-5);
            \draw[thick, green] (5,5,1)--(5,1,5);
            \draw[thick, green] (5,-1,5)--(5,-5,1);
            \draw[thick, dashed, green] (5,-5,-1)--(5,-1,-5);
            \draw[thick, dashed, green] (5,1,-5)--(5,5,-1);

            \draw[thick, blue] (0,1,0)--(-4,5,0);
            \draw[thick, blue] (0,-1,0)--(-4,-5,0);
            \draw[thick, blue] (0,0,1)--(-4,0,5);
            \draw[thick, dashed, blue] (0,0,-1)--(-4,0,-5);
            \draw[thick, blue] (-10,1,0)--(-6,5,0);
            \draw[thick, blue] (-10,-1,0)--(-6,-5,0);
            \draw[thick, blue] (-10,0,1)--(-6,0,5);
            \draw[thick, dashed, blue] (-10,0,-1)--(-6,0,-5);
            \draw[thick, blue] (-5,5,1)--(-5,1,5);
            \draw[thick, blue] (-5,-1,5)--(-5,-5,1);
            \draw[thick, dashed, blue] (-5,-5,-1)--(-5,-1,-5);
            \draw[thick, dashed, blue] (-5,1,-5)--(-5,5,-1);            
        \end{tikzpicture}
        \caption{}
        \label{subfig:redcheckatgreen}
    \end{subfigure}
    \hspace{0.02\textwidth} 
    \begin{subfigure}[t]{0.48\textwidth}
        \centering
        \begin{tikzpicture}[scale=0.37]
            \draw[thick, red] (-4,5,0)--(-5,5,1);
            \draw[thick, red] (0,1,0)--(0,0,1);
            \draw[thick, red] (-4,0,5)--(-5,1,5);
            \draw[red] (-4,5,0) circle(0.3);
            \draw[red] (-5,5,1) circle(0.3);
            \draw[red] (0,1,0) circle(0.3);
            \draw[red] (0,0,1) circle(0.3);
            \draw[red] (-4,0,5) circle(0.3);
            \draw[red] (-5,1,5) circle(0.3);

            \draw[thick, yellow] (-11,0,0)--(11,0,0);
            \foreach \i in {-5, 5}
                \draw[thick, yellow] ({\i},-6,0)--({\i},6,0);
            \foreach \i in {-5, 5}
                \draw[thick, yellow] ({\i},0,-6)--({\i},0,6);

            \draw[thick, green] (0,1,0)--(4,5,0);
            \draw[thick, green] (0,-1,0)--(4,-5,0);
            \draw[thick, green] (0,0,1)--(4,0,5);
            \draw[thick, dashed, green] (0,0,-1)--(4,0,-5);
            \draw[thick, green] (10,1,0)--(6,5,0);
            \draw[thick, green] (10,-1,0)--(6,-5,0);
            \draw[thick, green] (10,0,1)--(6,0,5);
            \draw[thick, dashed, green] (10,0,-1)--(6,0,-5);
            \draw[thick, green] (5,5,1)--(5,1,5);
            \draw[thick, green] (5,-1,5)--(5,-5,1);
            \draw[thick, dashed, green] (5,-5,-1)--(5,-1,-5);
            \draw[thick, dashed, green] (5,1,-5)--(5,5,-1);

            \draw[thick, blue] (0,1,0)--(-4,5,0);
            \draw[thick, blue] (0,-1,0)--(-4,-5,0);
            \draw[thick, blue] (0,0,1)--(-4,0,5);
            \draw[thick, dashed, blue] (0,0,-1)--(-4,0,-5);
            \draw[thick, blue] (-10,1,0)--(-6,5,0);
            \draw[thick, blue] (-10,-1,0)--(-6,-5,0);
            \draw[thick, blue] (-10,0,1)--(-6,0,5);
            \draw[thick, dashed, blue] (-10,0,-1)--(-6,0,-5);
            \draw[thick, blue] (-5,5,1)--(-5,1,5);
            \draw[thick, blue] (-5,-1,5)--(-5,-5,1);
            \draw[thick, dashed, blue] (-5,-5,-1)--(-5,-1,-5);
            \draw[thick, dashed, blue] (-5,1,-5)--(-5,5,-1);            
        \end{tikzpicture}
        \caption{}
        \label{subfig:redcheckatblue}
    \end{subfigure}

    \caption{This diagram shows two triangular operators. Figure~\ref{subfig:redcheckatgreen} illustrates an operator that is the product of checks along a loop on the green octahedron, marked by red circles. As expected, there are eight such operators on the green octahedron. We refer to this operator as a triangular-green operator. Similarly, Figure~\ref{subfig:redcheckatblue} shows the triangular-blue operator. These triangular operators commute with all elements of the SSG, as well as the blue and green checks, but they do not commute with the red or black checks and do not always mutually commute.}\label{fig:triangularoperator}
\end{figure}
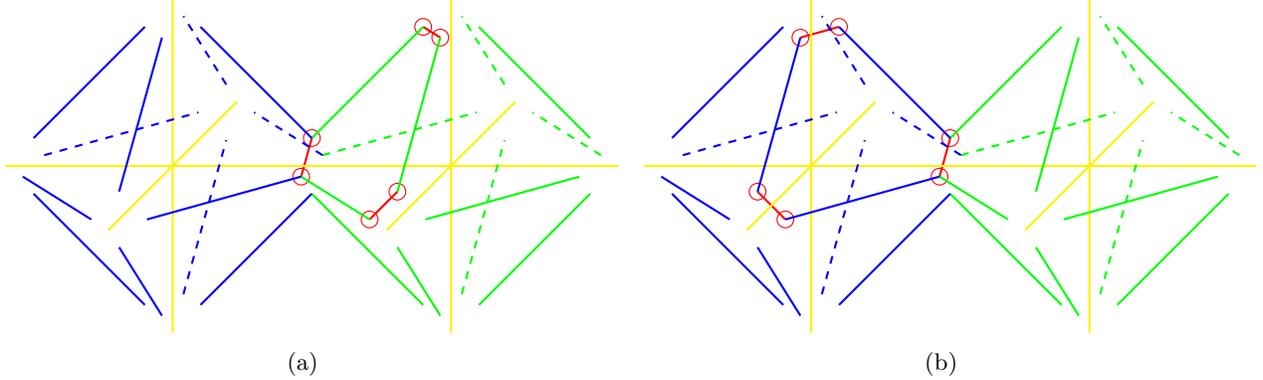

The measurement routine and ISG at each round is given by table~\ref{tab:3dISG}. Green/Blue stabilizers $W_g/W_b$ are the product of green/blue and red checks bordering the octahedron centered on the green/blue vertex. Red stabilizers are still operators formed by the product of inner-chain checks within each closed spin chain. We observe that only these three color operators commute with all checks and survive throughout the measurement routine, and thus we refer to them as the Steady Stabilizer Group (SSG). However, the Instantaneous Stabilizer Group (ISG) at round $r$ is given by $\text{ISG}_r = \text{SSG} \cup r\text{-checks} \cup \text{NSSG}_r$, where the Non-Steady Stabilizer Group (NSSG$_r$) contains instantaneous stabilizers whose values are randomized at certain rounds. We did not show explicit NSSG$_r$ as they will depend on the choice of the black check.

In Appendix~\ref{app:effectiveH}, we show that the SSG are equivalent to the stabilizers of the 3D toric code at step $6r + 5$, using the language of the 3D Kitaev spin liquid model.

The 3D toric code Hamiltonian, defined as the negative summation of all its stabilizers, follows the definition of the (3,1) toric code in \cite{freedman_double_2016}. One qubit is placed on each edge:
\begin{equation}
    H = -\sum_{v \in V} A_v - \sum_{p \in P} B_p,
\end{equation}\label{eqn:3dtoric}
The vertex term $A_v$ is defined as the application of the Pauli operator $X$ over the six edges connected to the vertex $v$, while $B_p$ refers to the application of the Pauli operator $Z$ over the four edges that form the boundary of the plaquette $p$. It is straightforward to see that these newly defined operators satisfy the relations $A_v^2 = B_p^2 = 1$ and commute with each other, i.e., $[A_v, B_p] = 0$.

\subsection{Error Correction}\label{subsec:3d error correction}
The 3D Floquet error correction is similar to the 2D case. We still label Pauli errors by color, corresponding to the action of the check operators (e.g., $X/Y/Z$ errors are labeled green / blue / red). There are four qubits placed on an edge (since four closed spin chains intersect at that edge), and six stabilizer operators have nontrivial action on these four qubits. These six stabilizers belong to the Steady Stabilizer Group (SSG), and are periodically updated, as listed in Table~\ref{tab:3d_floquet_routine}, making them useful for error detection.

Now, suppose that a green error occurs. One of the red stabilizers and the green stabilizer will flip. The flipped red stabilizer shares only one common edge with the green stabilizer, and this edge is, naturally, green. It is not surprising that the green errors at either end of the green check are indistinguishable. However, similar to the 2D case, we can apply a green operator to either end of the check, and it will correct the error regardless.

The blue error behaves similarly: it flips one red stabilizer and one blue stabilizer. We can apply a blue operator to either end of the common blue edge shared by the two flipped stabilizers to correct the error.

\begin{table}[htbp]
    \centering
    \begin{tabular}{|c|c|c|c|c|c|c|}
        \textbf{steps} & \textbf{6 r} & \textbf{6 r+1} & \textbf{6 r +2} & \textbf{6 r +3} & \textbf{6 r +4} & \textbf{6 r +5} \\
        Measure checks & Red and Black & Green & Blue & Red & Green & Blue  \\

        Updated SSG & Green & Blue & Red & --- & --- & red\\

    \end{tabular}
    \caption{The elements of the SSG are the product of check operators. The green/blue stabilizer is the product of the blue/green, red, and black checks on the octahedron centered at the green/blue vertex of $\Gamma$. The red stabilizer is the product of inner-chain checks along each closed spin chain. This table shows the operators in the SSG, whose values are updated immediately after each round of measurement.}\label{tab:3d_floquet_routine}
\end{table}

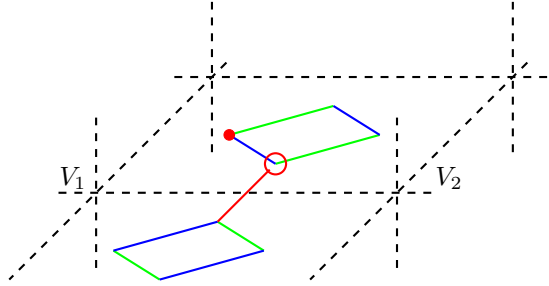
\begin{figure}[htbp]
    \centering
\begin{tikzpicture}[scale=1]
\foreach \i in {-2, 2}
        \draw[thick, dashed, black] ({\i},0,-4.5) -- ({\i},0,3);
\foreach \k in {-4, 0}
        \draw[thick, dashed, black] (-2.5,0,{\k}) -- (2.5,0,{\k});

\foreach \i in {-2, 2}
\foreach \k in {-4, 0}
        \draw[thick, dashed, black] ({\i},-1,{\k}) -- ({\i},1,{\k});

\draw[thick, green] (-1,0,-2) -- (0,0,-3);
\draw[thick, green] (0,0,-1) -- (1,0,-2);
\draw[thick, blue] (0,0,-3) -- (1,0,-2);
\draw[thick, blue] (-1,0,-2) -- (0,0,-1);

\draw[thick, green] (-1,0,2) -- (0,0,3);
\draw[thick, green] (0,0,1) -- (1,0,2);
\draw[thick, blue] (0,0,3) -- (1,0,2);
\draw[thick, blue] (-1,0,2) -- (0,0,1);

\filldraw[red] (-1,0,-2) circle (2pt);
\draw[thick, red] (0,0,1) -- (0,0,-0.8);
\draw[thick, red] (0,0,-1) circle (4pt);

\node[black] at (-2.5,0,-0.5) {$V_1$};
\node[black] at (2.5,0,-0.5) {$V_2$};
\end{tikzpicture}
\caption{An example of a red error: it will flip the octahedron operators centered on the vertices labeled $V_1$ and $V_2$. Conversely, when these two octahedron operators are flipped, we know that a red error has occurred. As before, we can apply a red operator to either end of the red check shown, which will either cancel the error or form a red or black check that will disappear automatically.}\label{fig:3derrorfigure}
\end{figure}
In 2D, all three colors are symmetric because the lattice is trivalent. However, even in the simplest 3D cubic cases, while the blue, red, and green checks still form a 3D trivalent lattice, the black checks break this symmetry in 3D. When a red error occurs, only the blue and green stabilizers are flipped. The common support for these two operators consists of four qubits on the same edge. We can apply another red error to any qubit around the same edge, which will either cancel the error or become part of the group generated by the red and black check operators, disappearing at step $6r + 5$. Thus, the entire code is proven to correct any single Pauli error.

One may ask why we don't use three colors and create a 3-step routine, given that error correction works. This leads us to consider logical information, which must commute with all stabilizers and remain invariant throughout the Floquet routine. The product of check operators along loops forms the inner logical information, while the outer logical operators are effectively the string operators on non-trivial loops on the 3-torus at step $6r + 5$. While the 3-step Floquet routine can correct errors, the logical operator collapses.

In the cubic lattice case, the red checks can be chosen so that the construction can be viewed as coupling layers, where the blue, green, and red checks form separate 2D toric code layers, and the black checks couple these layers into a 3D toric code. The string logical operator of the 3D toric code is the 2D toric code's electric logical operator, which commutes with the black checks. However, if a 3-step routine is applied, the 2D electric operator becomes a magnetic logical operator, which is randomized when the black checks are measured, as it does not commute with the black checks.

To address this, we adopt the "rewinding" technique from \cite{davydova_floquet_2023}, where doubling the measurement of blue, green, and red checks eventually maps the 2D electric logical operator back to itself, allowing it to survive the entire measurement routine. Details are provided in Appendix~\ref{app:logicevolution}.

\subsection{Criteria for the Error Correction of Floquet Codes}\label{subsec:IStopological}

It is common to require each instantaneous phase to be topological so that it can be framed within the automorphism codes framework \cite{aasen_adiabatic_2022}. However, we argue that this does not directly lead to an error-correctable Floquet code. First, having an instantaneous topological phase means that the elements in the $\text{ISG}$ can detect errors as syndrome operators.

Consider a Floquet code where three types of checks, $X \otimes X$, $Y \otimes Y$, and $Z \otimes Z$, are measured at specific steps, and $\text{ISG}$ is always topological. For simplicity, we assume that all the measurement outcomes of the checks are $+1$.

Suppose $X \otimes X$ is measured at round $r$. This implies that the two qubits connected by $X \otimes X$ effectively become one qubit. The $\text{ISG}_r = \text{SSG} \cup r\text{-checks} \cup \text{NSSG}_r$. The fact that the instantaneous phase at round $r$ is topological means that $\text{SSG} \cup \text{NSSG}_r$ forms a topological phase on the effective qubits. Thus, any effective single-qubit error, such as $X_{\text{eff}} = X \otimes 1 = 1 \otimes X$, $Z_{\text{eff}} = Z \otimes Z$, and $Y_{\text{eff}} = Y \otimes Z = Z \otimes Y$, must be distinguishable by $\text{SSG} \cup \text{NSSG}_r$. Interestingly, $X_{\text{eff}}$ corresponds to a single-qubit error in the Floquet code. The topological nature at round $r$ ensures that a single-qubit $X$ error is detectable and therefore error-correctable. Although $1 \otimes X$ and $X \otimes 1$ are indistinguishable (as they represent the same effective operator), this poses no issue for error correction since these two form a check operator, as discussed in Section~\ref{subsec:3d error correction}.

Similarly, single $Y$ and $Z$ errors are distinguishable at rounds where $Y \otimes Y$ and $Z \otimes Z$ are measured. Thus, if all $\text{ISG}$s form topological phases, all single Pauli errors are detectable by elements in the $\text{ISG}$. However, for the Floquet code to function, these operators must be referenced at least twice during the measurement routine before they are removed from the $\text{ISG}$, as syndrome operators are not measured directly in Floquet codes.

A direct example can be seen at step $6r+3$ in Table~\ref{tab:3dISG}. If we treat the construction as coupling layers, with red checks within each layer and black checks coupling layers, green and blue plaquette operators within each 2D layer as in Section~\ref{fig:2dinteractiondiagram} will appear. The red, blue, and green plaquette operators together form the full stabilizer group for uncoupled 2D toric code layers, similar to what is described in \cite{zhang_x-cube_2022}. However, only the blue plaquette operator is referenced again at round $6r + 4$, while the green plaquette operator is not referenced again before being removed from the ISG. Thus, these NSSG operators cannot be used for error correction. A similar situation occurs with the triangular-green operators, which appear in round $6r+1$ and are removed after round $6r+3$.

In contrast, elements in $\text{SSG}$ are periodically referenced because they are not removed from $\text{ISG}$. Any change in the referred value of $\text{SSG}$ can be used to detect errors as expected. Therefore, it is clearer for $\text{SSG}$ to handle error correction, as discussed throughout this paper. The key observation is that it is sufficient if an effective $\sigma_i$ error, which is a single-qubit $\sigma_i$ error, can be distinguished by $\text{SSG}$ when $\sigma_i \otimes \sigma_i$ is measured, where $i \in \{x, y, z\}$. In other words, in each instantaneous phase, $\text{SSG}$ should behave as a classical error-correcting code.

\begin{figure}[htbp]
    \centering
    \begin{subfigure}[b]{0.3\textwidth}
        \centering
        \begin{tikzpicture}[scale=1.2]
            \draw[step=1cm,gray,very thin] (0,0) grid (3,3);
            \foreach \x in {0,1,2,3} {
                \foreach \y in {0,1,2,3} {
                    \filldraw (\x,\y) circle (2pt);
                }
            }
            \draw[red, very thick] (0,1) -- (1,1); 
            \draw[green, very thick] (2,0) -- (2,1); 
        \end{tikzpicture}
        \caption{}\label{subfig:BP1}
    \end{subfigure}
    \hfill
    \begin{subfigure}[b]{0.3\textwidth}
        \centering
        \begin{tikzpicture}[scale=1.2]
            \draw[step=1cm,gray,very thin] (0,0) grid (3,3);
            \foreach \x in {0,1,2,3} {
                \foreach \y in {0,1,2,3} {
                    \filldraw (\x,\y) circle (2pt);
                }
            }
            \draw[green, very thick] (0,1) -- (3,1); 
            \draw[green, very thick] (0,2) -- (3,2); 
            \draw[red, very thick] (1,0) -- (1,3); 
            \draw[red, very thick] (2,0) -- (2,3); 
        \end{tikzpicture}
        \caption{}\label{subfig:BP2}
    \end{subfigure}
    \hfill
    \begin{subfigure}[b]{0.3\textwidth}
        \centering
        \begin{tikzpicture}[scale=1.2]
            \draw[step=1cm,gray,very thin] (0,0) grid (3,3);
            \foreach \x in {0,1,2,3} {
                \foreach \y in {0,1,2,3} {
                    \filldraw (\x,\y) circle (2pt);
                }
            }
            \draw[red, very thick] (0,0) -- (0,3); 
            \draw[green, very thick] (0,0) -- (3,0); 
        \end{tikzpicture}
        \caption{}\label{subfig:BP3}
    \end{subfigure}

    \vspace{1cm}

    \begin{subfigure}[b]{0.45\textwidth}
        \centering
        \begin{tikzpicture}[scale=0.9]
            \draw[step=1cm,gray,very thin] (0,0) grid (3,3);
            \foreach \x in {0,1,2,3} {
                \foreach \y in {0,1,2,3} {
                    \filldraw (\x,\y) circle (2pt);
                }
            }
            \draw[red, very thick] (0,0) -- (3,0); 
            \draw[red, very thick] (0,1) -- (3,1); 
            \draw[red, very thick] (0,2) -- (3,2); 
            
            \draw[->, very thick] (3.5,1.5) -- (4.0,1.5); 
            
            \begin{scope}[xshift=4.5cm]
                \foreach \y in {0,1,2,3} {
                    \filldraw (0,\y) circle (2pt);
                }
                \draw[green, thick] (0,0) -- (0,3); 
            \end{scope}
        \end{tikzpicture}
        \caption{}\label{subfig:BP4}
    \end{subfigure}
    \hfill
    \begin{subfigure}[b]{0.45\textwidth}
        \centering
        \begin{tikzpicture}[scale=0.9]
            \draw[step=1cm,gray,very thin] (0,0) grid (3,3);
            \foreach \x in {0,1,2,3} {
                \foreach \y in {0,1,2,3} {
                    \filldraw (\x,\y) circle (2pt);
                }
            }
            \draw[green, very thick] (0,0) -- (0,3); 
            \draw[green, very thick] (1,0) -- (1,3); 
            \draw[green, very thick] (2,0) -- (2,3); 
            
            \draw[->, very thick] (3.5,1.5) -- (4.0,1.5); 
            
            \begin{scope}[xshift=4.5cm]
                \foreach \x in {0,1,2,3} {
                    \filldraw (\x,0) circle (2pt);
                }
                \draw[red, thick] (0,0) -- (3,0); 
            \end{scope}
        \end{tikzpicture}
        \caption{}\label{subfig:BP5}
    \end{subfigure}

    \caption{A simple illustration of the 2-step Floquet Bacon-Shor code is provided. In this example, the code is defined on a $3 \times 3$ square lattice. Figure~\ref{subfig:BP1} shows the lattice with one qubit placed on each vertex. Each horizontal edge is associated with a $Z \otimes Z$ check, shown in red, and each vertical edge is associated with an $X \otimes X$ check, shown in green. Figure~\ref{subfig:BP2} shows the elements of the Stabilizer Group. The first type is the tensor product of $Z$ operators on qubits along any two consecutive vertical lines (red example), and the second type is the tensor product of $X$ operators on qubits along any two consecutive horizontal lines (green example). Figure~\ref{subfig:BP3} shows the logical operators of the subsystem code. The tensor product of $Z$ operators along the red line serves as the logical $Z_L$ operator, and the tensor product of $X$ operators along the green line serves as the logical $X_L$ operator. Figure~\ref{subfig:BP4} shows that when all red checks are measured at round $2r$, the four qubits on each horizontal line effectively become a single qubit, and an effective $Z \otimes Z$ acts on adjacent effective qubits. Figure~\ref{subfig:BP5} shows the case at round $2r+1$, where all green checks are measured, resulting in a similar effect. Essentially, both form a repetition code on different bases. It is important to note that the only logical information retained in this Floquet code is the inner logical information. Outer logical information can be found in \cite{alam_dynamical_2024} by adjusting the measurement routine.}\label{fig:BP} 
    
\end{figure}
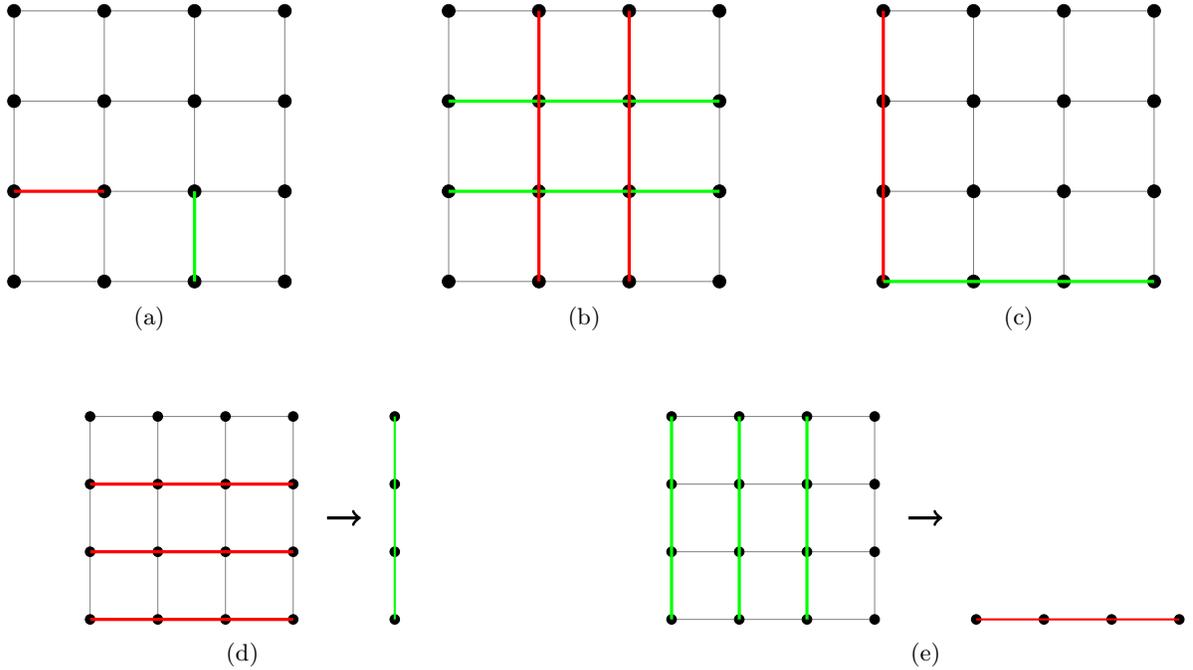

We find that the Floquet version of the Bacon-Shor code \cite{bacon_operator_2005} fits explicitly within this framework. For our purposes, we adopt a two-step measurement routine on a $L \times L$ square lattice, preserving only the inner logical operator, which corresponds to the logical operator of the parent Bacon-Shor subsystem code. In round $2r$, red checks are measured, and in round $2r+1$, green checks are measured. Each round yields an effective $L$-length repetition code, which is a classical error-correcting code. An explicit example is shown in Figure~\ref{fig:BP} on a $3 \times 3$ lattice. Thus, a well-aligned classical code may be used to construct an error-correctable Floquet code, which we leave for future work.

\subsection{Construction on General Lattices and Higher Dimensions}\label{subsec:construction on general lattices}
The coupling spin chain construction is purely localized, making it applicable to more general lattices. In 3D, if a lattice $\Gamma$ satisfies the following requirements:
1. All vertices are 2-colorable.
2. Each edge of $\Gamma$ borders an even number of plaquettes.

Then, a 3D error-correctable Floquet code can be defined. The assignment of colors is quite similar: we assign blue and green colors to the vertices, and the inner-chain checks are colored by the nearest vertex of $\Gamma$. For any edge $e$ bordering $N_q$ plaquettes, $N_q$ qubits are placed on the edge. The inter-spin chain checks are again separated into red and black. Half of the non-overlapping inter-chain checks ($\frac{N_q}{2}$) will be colored red, while the remaining ones will be black. This can be achieved under the second requirement above. The Floquet routine follows the same sequence as in Table~\ref{tab:3d_floquet_routine}, and the decoder is the same as in the 3D cubic case, using the $\text{SSG}$ as syndrome operators.

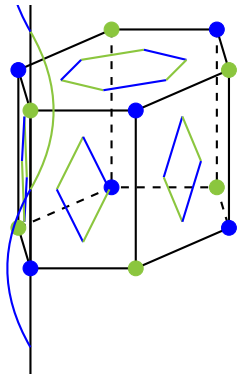
\begin{figure}[htbp]
    \centering
    \begin{tikzpicture}[scale=0.7]
    \clip (-2,-5) rectangle (4,2); 

    \draw[thick, black] (0,-7,0)--(0,6,0);
    \draw[thick, black] (-1,3,-2)--(0,3,0)--(2,3,0);
    \draw[thick, black] (-1,-6,-2)--(0,-6,0)--(2,-6,0);

    \draw[thick, black] (0,0,0)--(2,0,0)--(3,0,-2)--(2,0,-4)--(0,0,-4)--(-1,0,-2)--cycle; 
    \draw[thick, black] (-1,-3,-2)--(0,-3,0)--(2,-3,0)--(3,-3,-2); 
    \draw[thick, black] (-1,0,-2)--(-1,-3,-2); 
    \draw[thick, black] (0,0,0)--(0,-3,0); 
    \draw[thick, black] (2,0,0)--(2,-3,0); 
    \draw[thick, black] (3,0,-2)--(3,-3,-2); 
    \draw[thick, black, dashed] (0,0,-4)--(0,-3,-4); 
    \draw[thick, black, dashed] (2,0,-4)--(2,-3,-4);
    \draw[thick, black, dashed] (-1,-3,-2)--(0,-3,-4)--(2,-3,-4)--(3,-3,-2);

    \filldraw[LimeGreen] (0,0,0) circle (4pt);
    \filldraw[LimeGreen] (3,0,-2) circle (4pt);
    \filldraw[LimeGreen] (0,0,-4) circle (4pt);
    \filldraw[blue] (2,0,0) circle (4pt);
    \filldraw[blue] (-1,0,-2) circle (4pt);
    \filldraw[blue] (2,0,-4) circle (4pt);
    \filldraw[LimeGreen] (2,-3,0) circle (4pt);
    \filldraw[LimeGreen] (-1,-3,-2) circle (4pt);
    \filldraw[LimeGreen] (2,-3,-4) circle (4pt);
    \filldraw[blue] (0,-3,0) circle (4pt);
    \filldraw[blue] (3,-3,-2) circle (4pt);
    \filldraw[blue] (0,-3,-4) circle (4pt);

    \draw[thick, LimeGreen] (1,0,-1)--(0,0,-1.5);
    \draw[thick, blue] (1,0,-1)--(2,0,-1.5);
    \draw[thick, LimeGreen] (1,0,-3)--(0,0,-2.5);
    \draw[thick, blue] (1,0,-3)--(2,0,-2.5);
    \draw[thick, blue] (0,0,-1.5)--(0,0,-2.5);
    \draw[thick, LimeGreen] (2,0,-1.5)--(2,0,-2.5);

    \draw[thick, LimeGreen] (0.5,-1.5,0)--(1,-0.5,0);
    \draw[thick, blue] (0.5,-1.5,0)--(1,-2.5,0);
    \draw[thick, blue] (1.5,-1.5,0)--(1,-0.5,0);
    \draw[thick, LimeGreen] (1.5,-1.5,0)--(1,-2.5,0);

    \draw[thick, blue] (2.3,-1.5,-0.6)--(2.5,-0.5,-1);
    \draw[thick, LimeGreen] (2.3,-1.5,-0.6)--(2.5,-2.5,-1);
    \draw[thick, LimeGreen] (2.7,-1.5,-1.4)--(2.5,-0.5,-1);
    \draw[thick, blue] (2.7,-1.5,-1.4)--(2.5,-2.5,-1);

    \draw[thick, LimeGreen] (-0.3,-1.5,-0.6)--(-0.5,-0.5,-1);
    \draw[thick, blue] (-0.3,-1.5,-0.6)--(-0.5,-2.5,-1);
    \draw[thick, blue] (-0.7,-1.5,-1.4)--(-0.5,-0.5,-1);
    \draw[thick, LimeGreen] (-0.7,-1.5,-1.4)--(-0.5,-2.5,-1);

    \draw [thick, LimeGreen] (0,-1.5,0) to[out=60,in=-60] (0,1.5,0);
    \draw [thick, blue] (0,-1.5,0) to[out=-120,in=120] (0,-4.5,0);
    \draw [thick, blue] (0,1.5,0) to[out=120,in=-120] (0,4.5,0);
    \end{tikzpicture}
    \caption{The unit cell of a translationally invariant lattice with hexagonal symmetry does not support a coupling layer construction, as it does not satisfy the requirement that each edge must border an even number of plaquettes. However, extra closed spin chains can be introduced along each vertical axis, as shown. The checks of the vertical spin chains are colored similarly, enabling the construction of a valid error-correctable Floquet code.}
\label{fig:needvertical}
\end{figure}

An interesting example is a translationally invariant lattice composed of hexagonal prisms as its unit cell. This lattice is vertex-2-colorable, but the vertical edges border three plaquettes, which do not meet the above requirement. However, this can be compensated by introducing extra vertical spin chains, as shown in Figure~\ref{fig:needvertical}. New checks are assigned in the same way as usual checks, and an error-correctable Floquet code can be defined similarly. This further loosens the requirements for constructing a Floquet toric code to the following:
1. All vertices are 2-colorable.
2. Edges of $\Gamma$ that border an odd number of plaquettes form several closed loops.

Interestingly, since the current checks are 3-colored (except for the black checks), this naturally opens the possibility of a CSS Floquet code, which is a dynamical code that does not have a proper parent subsystem code, as explained in \cite{davydova_floquet_2023}. The checks are associated with different operators throughout the routine. In this case, the coupling spin chain construction only provides the coloring of the edges in the interaction diagram, without attaching fixed check operators. Therefore, we distinguish between "check" and "edge" here.

The measurement routine is shown in Table~\ref{tab:CSS_floquet_routine}. At each measurement step, the two-body operator of the specified type is measured on the two qubits at the ends of the given edge color. This arrangement yields a CSS Floquet code with preserved outer logical information, as described in \cite{davydova_floquet_2023}.

\begin{table}[htbp]
    \centering
    \begin{tabular}{|c|c|c|c|c|c|c|}
        \textbf{Steps} & \textbf{6r} & \textbf{6r+1} & \textbf{6r+2} & \textbf{6r+3} & \textbf{6r+4} & \textbf{6r+5} \\
        Edge Color & Blue and Red & Green & Red & Blue & Green & Red and Black \\
        Operator Type & Green & Red & Green & Red & Green & Red \\
    \end{tabular}
    \caption{Measurement routine for the CSS Floquet code. At each round, a two-body operator of the specified type is measured on the two qubits at the ends of the edges of the given color.}
    \label{tab:CSS_floquet_routine}
\end{table}

This generalization naturally extends to higher dimensions, using the same color arrangement for the edges, the same measurement routine shown in Table~\ref{tab:3d_floquet_routine}, and the same decoder introduced in Section~\ref{sec:3d construction}. The result is an $n$-dimensional error-correctable Floquet code with an instantaneous $n$-dimensional $(n,1)$ toric code phase, following the definition in \cite{freedman_double_2016}, where one qubit is placed on each 1-cell (the edge), and the Hamiltonian consists of the 0-cell terms (the vertex terms $A_v$) and the 2-cell terms (the plaquette terms $B_p$), similar to equation~\ref{eqn:3dtoric}:
\begin{equation}
    H = -\sum_{v \in V} A_v - \sum_{p \in P} B_p,
\end{equation}\label{eqn:ndtoric}
The vertex term $A_v$ is defined as the application of the Pauli operator $X$ over the all edges connected to the vertex $v$, while $B_p$ refers to the application of the Pauli operator $Z$ over the edges on the boundary of the plaquette $p$.

\subsection{$n$-Dimensional Floquet $X$-Cube Code}\label{subsec:Xcubefc}
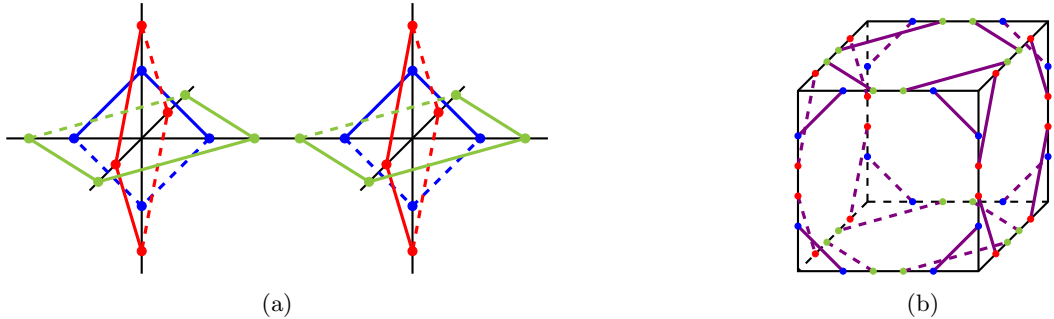
\begin{figure}[htbp]
    \centering
    \begin{subfigure}[b]
    {0.48\linewidth}
    
        \centering
        \begin{tikzpicture}[scale=0.6]
\foreach \i in {-3, 3}
        \draw[thick, black] ({\i},-3,0) -- ({\i},3,0);
\foreach \i in {-3, 3}
        \draw[thick, black] ({\i},0,-3) -- ({\i},0,3);
\draw[thick, black] (-6,0,0) -- (6,0,0);

\draw[very thick, blue] (1.5,0,0) -- (3,1.5,0) -- (4.5,0,0);
\draw[very thick, blue] (-1.5,0,0) -- (-3,1.5,0) -- (-4.5,0,0);
\draw[very thick, dashed, blue] (1.5,0,0) -- (3,-1.5,0) -- (4.5,0,0);
\draw[very thick, dashed, blue] (-1.5,0,0) -- (-3,-1.5,0) -- (-4.5,0,0);
\filldraw[blue] (1.5,0,0) circle (0.1);
\filldraw[blue] (3,1.5,0) circle (0.1);
\filldraw[blue] (4.5,0,0) circle (0.1);
\filldraw[blue] (3,-1.5,0) circle (0.1);
\filldraw[blue] (-1.5,0,0) circle (0.1);
\filldraw[blue] (-3,1.5,0) circle (0.1);
\filldraw[blue] (-4.5,0,0) circle (0.1);
\filldraw[blue] (-3,-1.5,0) circle (0.1);

\draw[very thick, red] (3,2.5,0) -- (3,0,1.5) -- (3,-2.5,0);
\draw[very thick, red] (-3,2.5,0) -- (-3,0,1.5) -- (-3,-2.5,0);
\draw[very thick, dashed, red] (3,2.5,0) -- (3,0,-1.5) -- (3,-2.5,0);
\draw[very thick, dashed, red] (-3,2.5,0) -- (-3,0,-1.5) -- (-3,-2.5,0);
\filldraw[red] (3,0,-1.5) circle (0.1);
\filldraw[red] (3,2.5,0) circle (0.1);
\filldraw[red] (3,0,1.5) circle (0.1);
\filldraw[red] (3,-2.5,0) circle (0.1);
\filldraw[red] (-3,0,-1.5) circle (0.1);
\filldraw[red] (-3,2.5,0) circle (0.1);
\filldraw[red] (-3,0,1.5) circle (0.1);
\filldraw[red] (-3,-2.5,0) circle (0.1);

\draw[very thick, LimeGreen] (0.5,0,0) -- (3,0,2.5) -- (5.5,0,0) -- (3,0,-2.5);
\draw[very thick, LimeGreen] (-5.5,0,0) -- (-3,0,2.5) -- (-0.5,0,0) -- (-3,0,-2.5);
\draw[very thick, dashed, LimeGreen] (0.5,0,0) -- (3,0,-2.5);
\draw[very thick, dashed, LimeGreen] (-5.5,0,0) -- (-3,0,-2.5);
\filldraw[LimeGreen] (0.5,0,0) circle (0.1);
\filldraw[LimeGreen] (5.5,0,0) circle (0.1);
\filldraw[LimeGreen] (3,0,2.5) circle (0.1);
\filldraw[LimeGreen] (3,0,-2.5) circle (0.1);
\filldraw[LimeGreen] (-0.5,0,0) circle (0.1);
\filldraw[LimeGreen] (-5.5,0,0) circle (0.1);
\filldraw[LimeGreen] (-3,0,2.5) circle (0.1);
\filldraw[LimeGreen] (-3,0,-2.5) circle (0.1);
\end{tikzpicture}
        \caption{}
        \label{subfig:3DXcubeconstruction}
    \end{subfigure}
    \hfill 
    \begin{subfigure}[b]{0.48\linewidth}
    
        \centering
        \begin{tikzpicture}[scale=0.4]
\draw[thick, dashed, black] (0,0,0) -- (6,0,0);
\draw[thick, dashed, black] (0,0,0) -- (0,6,0);
\draw[thick, dashed, black] (0,0,0) -- (0,0,6);
\draw[thick, black] (6,6,6) -- (6,6,0);
\draw[thick, black] (6,6,6) -- (0,6,6);
\draw[thick, black] (6,6,6) -- (6,0,6);
\draw[thick, black] (6,0,0) -- (6,6,0) -- (0,6,0) -- (0,6,6) -- (0,0,6) -- (6,0,6) -- cycle;

\draw[very thick, dashed, violet] (1.5,0,0) -- (0,1.5,0);
\draw[very thick, dashed, violet] (4.5,6,0) -- (6,4.5,0);
\draw[very thick, dashed, violet] (1.5,6,0) -- (0,4.5,0);
\draw[very thick, dashed, violet] (4.5,0,0) -- (6,1.5,0);
\draw[very thick, dashed, violet] (2.5,0,0) -- (0,0,2.5);
\draw[very thick, dashed, violet] (3.5,0,6) -- (6,0,3.5);
\draw[very thick, dashed, violet] (2.5,0,6) -- (0,0,3.5);
\draw[very thick, dashed, violet] (3.5,0,0) -- (6,0,2.5);
\draw[very thick, dashed, violet] (0,2.5,0) -- (0,0,1.5);
\draw[very thick, dashed, violet] (0,3.5,6) -- (0,6,4.5);
\draw[very thick, dashed, violet] (0,2.5,6) -- (0,0,4.5);
\draw[very thick, dashed, violet] (0,3.5,0) -- (0,6,1.5);
\draw[very thick, violet] (1.5,0,6) -- (0,1.5,6);
\draw[very thick, violet] (4.5,6,6) -- (6,4.5,6);
\draw[very thick, violet] (1.5,6,6) -- (0,4.5,6);
\draw[very thick, violet] (4.5,0,6) -- (6,1.5,6);
\draw[very thick, violet] (2.5,6,0) -- (0,6,2.5);
\draw[very thick, violet] (3.5,6,6) -- (6,6,3.5);
\draw[very thick, violet] (2.5,6,6) -- (0,6,3.5);
\draw[very thick, violet] (3.5,6,0) -- (6,6,2.5);
\draw[very thick, violet] (6,2.5,0) -- (6,0,1.5);
\draw[very thick, violet] (6,3.5,6) -- (6,6,4.5);
\draw[very thick, violet] (6,2.5,6) -- (6,0,4.5);
\draw[very thick, violet] (6,3.5,0) -- (6,6,1.5);

\foreach \i in {2.5, 3.5}
\foreach \j in {0, 6}
\foreach \k in {0, 6}
        \filldraw[LimeGreen] ({\i},{\j},{\k}) circle (0.1);
\foreach \i in {0, 6}
\foreach \j in {0, 6}
\foreach \k in {2.5, 3.5}
        \filldraw[LimeGreen] ({\i},{\j},{\k}) circle (0.1);
\foreach \i in {1.5, 4.5}
\foreach \j in {0, 6}
\foreach \k in {0, 6}
        \filldraw[blue] ({\i},{\j},{\k}) circle (0.1);
\foreach \i in {0, 6}
\foreach \j in {1.5, 4.5}
\foreach \k in {0, 6}
        \filldraw[blue] ({\i},{\j},{\k}) circle (0.1);
\foreach \i in {0, 6}
\foreach \j in {0, 6}
\foreach \k in {1.5, 4.5}
        \filldraw[red] ({\i},{\j},{\k}) circle (0.1);
\foreach \i in {0, 6}
\foreach \j in {2.5, 3.5}
\foreach \k in {0, 6}
        \filldraw[red] ({\i},{\j},{\k}) circle (0.1);
\end{tikzpicture}
        \caption{}
        \label{subfig:unit cell of X cube}
    \end{subfigure}
    \caption{Figure~\ref{subfig:3DXcubeconstruction} illustrates the placement of spin chains on a 3D cubic lattice. Spin chains of the same color are confined to the same planar slide. Within each planar slide, the coloring of checks follows the same scheme as in Figure~\ref{subfig:fcondual}. This setup generalizes the placement of spin chains on the faces of the dual lattice in 2D, placing three closed spin chains around each vertex of $\Gamma$. Each physical edge is crossed by four spin chains. Figure~\ref{subfig:unit cell of X cube} shows the unit cell of this construction. All purple-colored edges represent inner-chain checks, which are colored within each planar slide. The product of all checks within the unit cell gives the cubic stabilizers.}\label{fig:3dconstruction}
\end{figure}

In 2D, there are two ways to place closed spin chains, both leading to the same Floquet code, as discussed in Section~\ref{subsec:2dconstructions}. Generalizing this approach by placing spin chains on the faces of $\Gamma$ results in the $n$-dimensional toric code Floquet code. In this section, we extend this idea based on the diagram in Figure~\ref{fig:2ddual}. A 3D cubic lattice can be treated as intersecting transversal planar slices, as shown in Figure~\ref{fig:3dconstruction}. We place closed spin chains around the vertices in the planes $x$-$y$, $x$-$z$, and $y$-$z$, respectively. When the spin chains intersect at an edge, a $Z \otimes Z$ coupling is introduced to enable interaction between them.

Interestingly, the coloring of checks must be determined within each planar slice. On each 2D slice of the lattice, the plaquettes are 2-colored, and the inner-chain checks are assigned the color of the plaquette to which they belong, similar to the 2D coloring in Figure~\ref{subfig:fcondual}. Furthermore, there is no flexibility in separating red and black checks: the red checks must be the inter-chain checks within each 2D planar slice, while the remaining checks are colored black. Following this coloring scheme, we can apply the same measurement routine as outlined in Table~\ref{tab:3d_floquet_routine} to realize the 3D $X$-cube Floquet code.

The 3D $X$-cube code is defined on a 3D cubic lattice, with one qubit placed on each edge of the lattice:
\begin{equation}
    H=-\sum_{v \in V} (A_v^{x,y} + A_v^{y,z} + A_v^{x,z}) - \sum_{c \in C} B_c,
\end{equation}
where $V$ and $C$ represent the sets of vertices and cubes, respectively. The operator $A_v^{i,j}$, with $i, j \in \{x, y, z\}$, applies the Pauli $Z$ operator to the qubits on the four edges connected to vertex $v$ within the planar slice spanned by the $i$ and $j$ axes. The term $B_c$ applies the Pauli $X$ operator to the twelve edges within the cube $c$. These operators satisfy $(A_v^{i,j})^2 = B_c^2 = 1$ and $[A_v^{i,j}, B_c] = 0$.

The Steady Stabilizer Group (SSG) of this Floquet code consists of two types of terms. The first type is the vertex term, which is the product of inner-chain checks along each closed spin chain. The second type is the cubic term, which is the product of all checks within each unit cube, as shown in Figure~\ref{subfig:unit cell of X cube}. Error correction can again be performed using the SSG alone. Note that all closed spin chains are connected by inter-chain coupling checks, ensuring they never overlap. When a red error (Pauli $Z$) occurs on a qubit, four cubic terms will flip. The common support of these four cubic terms consists of the four qubits connected by inter-chain $Z \otimes Z$ couplings. If a green or blue error occurs, it will flip two cubic terms and one vertex term. The common support of these three operators involves a single check of the same color as the error. Thus, the error correction process is similar to that of the Floquet toric code.

At round $6r$, the vertex terms correspond to $A_v^{i,j}$, and the cubic terms correspond to $B_c$, as expected. This construction recovers the $X$-cube Floquet code in three dimensions \cite{zhang_x-cube_2022}, but with a different decoding scheme.

Moreover, this construction is purely localized, avoiding the need for global coupling layers, and can be easily applied to construct the $X$-cube Floquet code on any manifold, consistent with previous results \cite{shirley_fracton_2018}. More interestingly, on $n$-dimensional lattices composed of transversely intersecting surfaces, we can place spin chains around each vertex, with each spin chain lying in a 2D plane within the local cube. The coloring of edges follows the same rules as in the 3D case. As a result, we obtain an extended $X$-cube model on an $n$-dimensional lattice $\Gamma$ at round $6r$:

\begin{equation}
    H=-\sum_{v \in V}\sum_{x_i,x_j} A_v^{x_i,x_j} - \sum_{c \in C} B_c,
\end{equation}\label{eqn:higherXcube}
where $x_i$ indexes the spatial axes, and $A_v^{x_i,x_j}$ applies the Pauli $Z$ operator to the nearest qubits on the local 2D plane spanned by $x_i$ and $x_j$. Here, $c$ denotes the n-cell of $\Gamma$, and $B_c$ acts on the qubits within the n-cell $c$.

The extended model yields results similar to those of the 3D $X$-cube model, where the logarithm of the ground-state degeneracy scales as $\log(\text{GSD}) \approx L^{n-2}$ on an $n$-dimensional cubic lattice of length $L$. It is also a fracton model, exhibiting lineon and hyper-planon excitations. Further details can be found in Appendix~\ref{app:X}.

\section{Conclusion} \label{sec:conclusion}

In this paper, we demonstrate how the 2D Floquet code can be generally integrated into a coupling spin chain construction. We distinguish real lattice edges from the interaction diagrams composed of interaction checks. Under this construction, the previous requirement of a trivalent and 3-colorable check network is replaced by the simpler condition that the real lattice vertices are 2-colorable. By generalizing the placement of closed spin chains over the plaquettes of the lattice, we provide an explicit construction of an $n$-dimensional Floquet code with an instantaneous $n$-dimensional $\mathbf{Z}_2$ phase. Since the coupling spin chain construction is purely localized, our Floquet code can be applied to any dimensional lattice, as long as its vertices are 2-colorable.

We also describe how the Floquet state evolves and explain the emergence of an exact $n$-dimensional $(n,1)$ toric code. It is important to note that the $n$-dimensional toric code only appears instantaneously, and the Floquet routine does not obviously induce an automorphism of topological orders. However, we demonstrate that it is sufficient to use the Steady Stabilizer Group (SSG) for error correction, with logical information preserved in the 6-step measurement routine. Therefore, a topological Floquet code is error-correctable without requiring all instantaneous phases to be topological. As outlined in Section~\ref{subsec:IStopological}, the Floquet code remains error-correctable if the SSG forms an instantaneous classical error-correcting code at all times. We explicitly present the 2-step Floquet Bacon-Shor code, which holds instantaneous repetition codes at each round.

Additionally, the 3-coloring of checks provides a natural framework for constructing a CSS Floquet code when periodic check measurements are implemented. We argue that this CSS Floquet code is error-correctable and capable of carrying logical information, serving as a counterexample to the general assumption that coupling wires or spin chains always lead to fractonic phases, as suggested in \cite{ma_fracton_2017, fuji_coupled_2019, sullivan_fractonic_2021, williamson_type-ii_2021, halasz_fracton_2017, hsieh_fractons_2017}.

The $X$-cube Floquet code \cite{zhang_x-cube_2022} can be understood by placing closed spin chains around vertices within each planar slice, generalizing the placement of closed spin chains on the dual lattice plaquettes as seen in Figure~\ref{subfig:fcondual}. Error correction can also be performed using the Steady Stabilizer Group (SSG) alone.

Our localized construction allows us to explicitly visualize the foliation structure. Further analysis can extend the construction to any lattice that is locally cubic-like, aligning with the general 3D $X$-cube model on arbitrary manifolds \cite{shirley_fracton_2018}. This approach also extends naturally to higher dimensions, where we obtain an error-correctable Floquet code with instantaneous higher-dimensional $X$-cube behavior. Using the Laurent polynomial method on an $n$-dimensional hypercubic lattice of size $L$ with periodic boundary conditions, we show that the ground state degeneracy (GSD) satisfies:

\begin{equation}
    \log_2(\text{GSD}) = N_q - N_S = 2 \cdot C^2_n L^{n-2} + \text{poly}(L, n-3),
\end{equation}\label{eqn:higherXcubeGSD}

where $\text{poly}(L, n-3)$ is a polynomial of degree $n-3$. This model is topological, exhibiting lineon and hyper-planon excitations. Lineons, in particular, can move in extended dimensions as long as an $n-2$ dimensional multipole is paired.

In conclusion, the coupling spin chain construction localizes the traditional coupling layer approach, naturally providing a parent subsystem code. Two different families of Floquet codes can be constructed in dimensions higher than two and on more general lattices. Both are error-correctable and carry logical information. We argue that Floquet codes remain error-correctable when the SSG forms an instantaneous classical error-correcting code, as demonstrated by the Floquet Bacon-Shor code. This suggests that a well-aligned classical error-correcting code could provide a more general framework for constructing quantum Floquet codes, a topic we leave for future work.

We also identify a special type of subsystem code that behaves similarly to the 3D toric code when a maximal commuting set of gauge checks is added to the stabilizer group. This demonstrates the potential of the spin chain construction in discovering new topological phases. In the future, we aim to explore more topological phases that can be constructed using this approach and investigate how the coupling spin chain might offer new Floquet code possibilities. Additionally, CSS Floquet codes, although not typically associated with a useful parent code, might be described through anyon condensation of certain color codes. This is particularly interesting in our case, as it may lead to symmetry-broken higher-dimensional color codes.

\vspace{0.5cm}
\noindent\textbf{Acknowledgement.} 
The authors are partially supported by NSF grant CCF-2006667, Quantum Science Center sponsored by DOE's Office of Science, and ARO MURI. 

\bibliographystyle{plain}
\bibliography{references}

\onecolumn\newpage
\appendix
\section{The Instantaneous Phase of Floquet code in dimension 3}\label{app:effectiveH}
Consider a $3$-dimensional Kitaev Spin Liquid(KSL) Hamiltonian:
    \begin{equation}
        H=\sum - J_x X \otimes X - J_Y Y \otimes Y- J_z Z \otimes Z
    \end{equation}
The arrangement of check operators is shown in figure~\ref{subfig:3dtoriccube}. Green/Blue checks are associated with $X \otimes X$/$Y \otimes Y$ operators and controlled by $J_x$/$J_y$, and red and black checks are all $Z\otimes Z$ operators, controlled by $J_z$. The check operators no longer always anti-commute when they are connected.
When $J_z$ is dominant, we will get the instantaneous phase at round $6r$. Each group of 4 qubits near each physical edge will falls into the $+1$ common eigenspace of operators $Z \otimes Z \otimes I \otimes I$, $I \otimes Z \otimes Z \otimes I$ and $I \otimes I \otimes Z \otimes Z$, thus become effective one qubit. 
Effective Pauli operators are generated by $\mathbf{Z} = Z \otimes I \otimes \dots \otimes I$ and $\mathbf{X} = X \otimes X \otimes X \otimes X$. It is easy to see, the red stabilizers in SSG are products of $\mathbf{X}$ along the border of each plaquette and the blue/green stabilizers are products of $\mathbf{X}$ over each edges connected to the vertex. We shall note here, if one edge only borders odd number of qubits, that two vertex opertors that the edge connects will anticommute. So, the requirement of possible splitting of red and black checks agree with the commutativity of effective stabilizer operators. The above argument only require the whole lattice is vertex-2-colorable and each edge borders even number of plaquettes, which matchs the requirement of construcing the floquet code and works on general lattices. So it naturally works also in higher dimensions, thus will generate $(n,1)$ toric code model at round $6r$. 

For the $X$-cube construction, a unit cell of cubic lattice as shown in figure~\ref{subfig:3DXcubeconstruction}, edges of the same color represents the edges from the same spin chain, and $Z\otimes Z$ inter-chain coupling are in between concussive qubits on the same edge. Effective operators are cubic terms, that are product of all check operators shown in this figure, and the vertices terms, that are the product of check operators of the same color around a single vertex, noticing that the arrangement of edge operators are translational-invariant. Each edge is surrounded by $4$ cube terms, and the action of them on this edge are $X \otimes X \otimes X \otimes X$, $X \otimes Y \otimes X \otimes Y$, $Y \otimes Y \otimes Y \otimes Y$, $Y \otimes X \otimes Y \otimes X$, respectively. They commute so they have the same effective representation $\mathbf{X}$. The action of vertex terms on each edge is the effective operator $\mathbf{Z}$. So they matches the Hamiltonian of X Cube model. The analysis works naturally to higher dimensions, thus permits $n$-dimensional instantaneous (4,1) toric code and generalized $X$-cube model.

\section{A Trivalent 3D Kitaev Spin Liquid Model from Coupling Spin Chain}
\subsection{A Short Review of Binary Vector Representation of the Pauli Hamiltonian}

Single-qubit Pauli operators, or the Pauli matrices, are denoted by $X$, $Y$, $Z$, or $\sigma_x$, $\sigma_y$, $\sigma_z$, satisfying
\[
\sigma_i \cdot \sigma_j = \delta_{i,j} + i\epsilon_{ijk} \sigma_k.
\]
Thus, the tensor product of Pauli matrices over finite support forms an abelian group $P$, with scalars of $\{\pm 1, \pm i\}$. When considering the stabilizer code, one can ignore the scalar coefficients, leading to the abelian Pauli group $P/\{\pm 1, \pm i\}$. Thus, we can define an $\mathbf{F}_2$ module vector representation for Pauli matrices as follows:
\[
\text{Pauli-X} \mapsto \begin{bmatrix} 1 \\ 0 \end{bmatrix}, \quad
\text{Pauli-Y} \mapsto \begin{bmatrix} 1 \\ 1 \end{bmatrix}, \quad
\text{Pauli-Z} \mapsto \begin{bmatrix} 0 \\ 1 \end{bmatrix}.
\]
To restate the commutation relation in this new language, we shall use the symplectic matrix.

\[
\lambda = 
\begin{bmatrix}
0 & 1 \\
1 & 0
\end{bmatrix}
\]
We call the map $\tau: P \to V$ as $V_P$, where $P$ is a Pauli operator, and $V_P$ is the corresponding vector representation. The commutation relation $[P, P'] = 0$ is restated as $V_P^T \cdot \lambda \cdot V_P' = 0 \ (\text{mod}\ 2)$. Here, $T$ denotes transpose.

For a Pauli operator that is the tensor product of $K$ Pauli matrices, we can use a $2K$-dimensional vector to represent it, i.e., the direct sum of the vector representation of each Pauli matrix. We rearrange the vector so that the first (last) $K$ entries mainly record the $\sigma_x$ ($\sigma_z$) on $K$ sites.

For a translationally invariant stabilizer code, one can write the Hamiltonian term in a compact form. An $n$-dimensional translationally invariant lattice is denoted as a binary polynomial ring $\mathbf{F}_2[x_1, x_2, \dots, x_n]$. Suppose each unit cell of the lattice contains $q$ qubits. Then a generator of a translationally invariant stabilizer code can be denoted as:
\[
\text{P} \mapsto \begin{bmatrix} L_1(x_1, x_2, \dots, x_n) & L_2(x_1, x_2, \dots, x_n) & \dots & L_{2q}(x_1, x_2, \dots, x_n) \end{bmatrix}
\]
where $L_i(x_1, x_2, \dots, x_n)$ is a Laurent polynomial over $\mathbf{F}_2$. 

A simple but non-trivial example is to consider a 2D lattice, denoted as $\mathbf{F}_2[x, y]$, with 2 qubits placed on each vertex, labeled by $a$ and $b$, respectively. The Hamiltonian, which is the negative sum of all stabilizers, is written as:
\[
H =  - \sum_{i,j} X^a_{i,j} \otimes X^a_{i+1,j} \otimes X^b_{i,j+1}.
\]
The subscript $\{i,j\}$ denotes the position of the qubit being acted upon. We can pick any point on the lattice as the origin and denote this origin as position 1. The Hamiltonian term at the origin is referred to as the generator, since all other terms can be obtained by translating this generator. The generator applies $\sigma_x$ to the first qubit of the $a$-labeled qubit at position $1$ and $x$, and applies $\sigma_x$ to the $b$-labeled qubit at position $y$. It can be written as the following binary vector:

\[
h =\begin{bmatrix} 1+x & y & 0 & 0 \end{bmatrix}.
\]
All Hamiltonian terms can be generated by the formula $M(x, y) \cdot h$, where $M(x, y)$ is a monomial of $x$ and $y$, representing the position of the Hamiltonian term.

If there are more Hamiltonian generators, we simply have more vectors. Thus, a Hamiltonian or a stabilizer code can be written as a $k \times 2q$ matrix, where $k$ is the number of generators, and each entry of the matrix is a Laurent polynomial. For a more detailed reference, see\cite{haah_lattice_2013}.

\subsection{The Ground State Degeneracy of the 3D Trivalent Model}

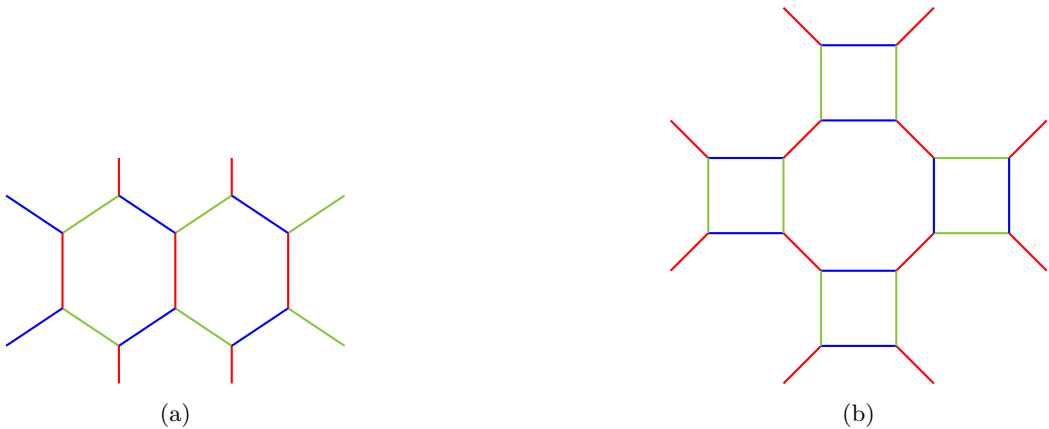
\begin{figure}[htbp]
    \centering
    
    \begin{subfigure}[b]{0.45\textwidth}
        \centering
        \begin{tikzpicture}[scale=0.5]
            \foreach \i in {-3,0,3}
            \foreach \j in {-1,1}
                    \draw[thick, LimeGreen] ({\i},{\j}) -- ({\i+1.5},{\j*2});
            \foreach \i in {-3,0,3}
            \foreach \j in {-1,1}
                    \draw[thick, blue] ({\i},{\j}) -- ({\i-1.5},{\j*2});
            \foreach \i in {-3,0,3}
                    \draw[thick, red] ({\i},-1) -- ({\i},1);
            \foreach \i in {-1.5,1.5}
            \foreach \j in {-3,2}
                    \draw[thick, red] ({\i},{\j}) -- ({\i},{\j+1});
        \end{tikzpicture}
        \caption{}
        \label{subfig:honeycombcouple}
    \end{subfigure}
    \hfill
    \begin{subfigure}[b]{0.45\textwidth}
        \centering
        \begin{tikzpicture}[scale=0.5]
            \foreach \i in {-1,1}
            \foreach \j in {-1,1}
                    \draw[thick, red] ({\i*2},{\j}) -- ({\i},{\j*2});
            \foreach \i in {-1,1}
            \foreach \j in {-1,1}
                    \draw[thick, red] ({\i*4},{\j}) -- ({\i*5},{\j*2});
            \foreach \i in {-1,1}
            \foreach \j in {-1,1}
                    \draw[thick, red] ({\i},{\j*4}) -- ({\i*2},{\j*5});
            \foreach \j in {-4,-2,2,4}
                    \draw[thick, blue] (-1,{\j}) -- (1,{\j});
            \foreach \i in {-1,1}
            \foreach \j in {-4,2}
                    \draw[thick, LimeGreen] ({\i},{\j}) -- ({\i},{\j+2});
            \foreach \j in {-1,1}
                    \draw[thick, blue] (-4,{\j}) -- (-2,{\j}); 
            \foreach \i in {-2,-4}
                    \draw[thick, LimeGreen] ({\i},-1) -- ({\i},1);         
            \foreach \j in {-1,1}
                    \draw[thick, LimeGreen] (2,{\j}) -- (4,{\j}); 
            \foreach \i in {2,4}
                    \draw[thick, blue] ({\i},-1) -- ({\i},1);
        \end{tikzpicture}
        \caption{}
        \label{subfig:488couple}
    \end{subfigure}

    \caption{\ref{subfig:honeycombcouple} places spin chains extending horizontally and couples them vertically in a specific manner. This setup recovers the honeycomb Kitaev spin liquid model and supports the toric code phase in the limit of strong inter-chain coupling. However, note that this is not a natural Floquet code construction (unless the color of the checks is rearranged). \ref{subfig:488couple} places closed spin chains and couples them as a square-octagon lattice. Similarly, this can be viewed as a Kitaev spin liquid model, and a $\mathbf{Z}_2$ phase is recovered at the strong inter-chain coupling limit. Indeed, there are various ways to place the spin chains, and they will always recover the $\mathbf{Z}_2$ phase in the strong coupling limit when the interaction diagram is trivalent.}
    \label{fig:overall}
\end{figure}

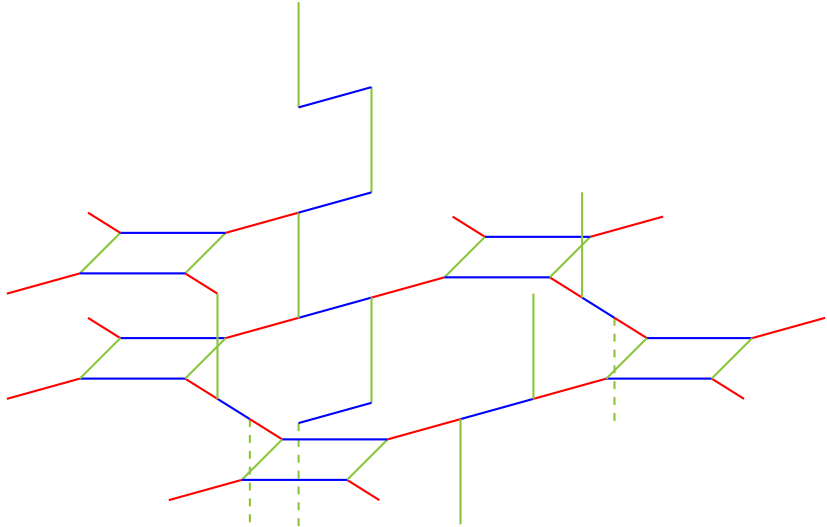
\begin{figure}[htbp]
    \centering
    \begin{tikzpicture}[scale=0.7]
\draw[thick, LimeGreen, dashed] (-3,-2,-2) -- (-3,-4,-2);
\draw[thick, LimeGreen, dashed] (-2,0,3) -- (-2,-2,3); 
\draw[thick, LimeGreen, dashed] (3,0,-2) -- (3,-2,-2); 

\foreach \i in {-1, 1}
\foreach \k in {-1, 1}
        \draw[thick, blue] ({\i*2},0,{\k*3}) -- ({\i*3},0,{\k*2});
\foreach \i in {-1, 1}
\foreach \k in {-1, 1}
        \draw[thick, red] ({\i*4},0,{\k*1}) -- ({\i*3},0,{\k*2});
\foreach \i in {-1, 1}
\foreach \k in {-1, 1}
        \draw[thick, red] ({\i*2},0,{\k*3}) -- ({\i*1},0,{\k*4});
\foreach \i in {-1, 1}
\foreach \k in {-1, 1}
        \draw[thick, red] ({\i*6},0,{\k*1}) -- ({\i*7},0,{\k*2});
\foreach \i in {-1, 1}
\foreach \k in {-1, 1}
        \draw[thick, red] ({\i*1},0,{\k*6}) -- ({\i*2},0,{\k*7});
\foreach \i in {-1, 1}
\foreach \k in {-1, 1}
        \draw[thick, LimeGreen] ({\i*1},0,{\k*4}) -- ({\i*1},0,{\k*6});
\foreach \i in {-1, 1}
\foreach \k in {-1, 1}
        \draw[thick, blue] ({\i*4},0,{\k*1}) -- ({\i*6},0,{\k*1});
\foreach \i in {-6,-4,4,6}
        \draw[thick, LimeGreen] ({\i},0,-1) -- ({\i},0,1);
\foreach \k in {-6,-4,4,6}
        \draw[thick, blue] (-1,0,{\k}) -- (1,0,{\k});

\draw[thick, LimeGreen] (-4,2,-1) -- (-4,2,1);
\draw[thick, LimeGreen] (-6,2,-1) -- (-6,2,1);
\draw[thick, blue] (-4,2,-1) -- (-6,2,-1);
\draw[thick, blue] (-4,2,1) -- (-6,2,1);
\draw[thick, red] (-4,2,1) -- (-3,2,2);
\draw[thick, red] (-4,2,-1) -- (-3,2,-2);
\draw[thick, red] (-6,2,1) -- (-7,2,2);
\draw[thick, red] (-6,2,-1) -- (-7,2,-2);
\draw[thick, LimeGreen] (-3,0,-2) -- (-3,2,-2);
\draw[thick, LimeGreen] (-3,4,-2) -- (-3,6,-2);
\draw[thick, LimeGreen] (-2,2,-3) -- (-2,4,-3);
\draw[thick, LimeGreen] (-2,0,-3) -- (-2,-2,-3); 
\foreach \j in {-2, 2, 4}
        \draw[thick, blue] (-3,{\j},-2) -- (-2,{\j},-3);
\draw[thick, LimeGreen] (-3,0,2) -- (-3,2,2); 

\draw[thick, LimeGreen] (2,0,-3) -- (2,2,-3); 
\draw[thick, LimeGreen] (3,0,2) -- (3,2,2); 
\draw[thick, LimeGreen] (2,0,3) -- (2,-2,3); 
\end{tikzpicture}
    \caption{It shows a special spin chain construction of a 3d Kitaev spin liquid model. Closed spin chains on squares are places on each plane and coupled to vertically spin chains . We prove it is topological phase that share the same GSD with 3d toric code at the strong coupling limit. It also have the mobile point like charges excitations. }
    \label{fig:special3d}
\end{figure}

In 2D, different placements of spin chains result in the same phase, as shown in Figure~\ref{subfig:488couple} and Figure~\ref{subfig:honeycombcouple}, consistent with the fact that a 2D translationally invariant Pauli Hamiltonian can only support a $\mathbf{Z}_2$ topological phase. However, in 3D, there are many more distinct topological and exotic phases. Here, we identify a special coupling spin chain construction, shown in Figure~\ref{fig:special3d}. This construction deviates significantly from the traditional coupling layer approach, but we prove that it has the same ground state degeneracy (GSD) as the 3D toric code on any manifold, though it is unknown whether they are fully equivalent. This suggests that coupling spin chain constructions could recover more topological phases or even lead to new Floquet codes.

In the context of the Kitaev spin liquid (KSL) or equivalently a gauge-fixed subsystem code, the effective Hamiltonian can be calculated using perturbation theory \cite{kitaev_anyons_2008}\cite{yan_generalized_2024}. To rigorously demonstrate that this special construction shares the same GSD as the 3D toric code, we adopt the Laurent polynomial representation of the effective Hamiltonian.

\[
H =
\left(
\begin{array}{cccccccccccc}
 x+1 & y+1 & 0 & 0 & 0 & 0 & 0 & 0 & 1 & 1 & 1 & 1 \\
 0 & 0 & 0 & 0 & 0 & 0 & y+1 & x y+y & y & x y & x & 1 \\
 0 & z+1 & z+1 & z+1 & 0 & 0 & 0 & 0 & z & z & 0 & 0 \\
 0 & y z+y & 0 & 0 & z+1 & z+1 & 0 & 0 & 0 & 0 & z & z \\
 z+1 & 0 & 0 & z+1 & z+1 & 0 & 0 & 0 & 0 & 1 & 1 & 0 \\
 0 & 0 & 0 & 0 & x z+x & z+1 & z+1 & 0 & 0 & 0 & x & 1 \\
 0 & 0 & z+1 & x z+x & 0 & 0 & z+1 & 0 & 1 & x & 0 & 0 \\
 0 & 0 & y z+y & 0 & 0 & z+1 & 0 & y z+y & y z & 0 & 0 & z \\
\end{array}
\right)
\]
Each column represents a generator of the Hamiltonian. Thus, the Hamiltonian contains 8 generators, and each unit cell has 6 qubits. We find the following invertible matrices:

\[
r = 
\left(
\begin{array}{cccccccccccc}
 0 & 0 & 0 & 0 & 0 & 0 & 0 & 0 & 0 & 0 & 0 & 1 \\
 0 & 0 & 0 & 0 & 0 & 0 & 0 & 0 & 1 & 0 & 0 & 1 \\
 0 & 0 & 0 & 0 & 0 & 0 & 0 & 0 & 0 & 0 & 1 & 1 \\
 0 & 0 & 0 & \bar{x} & \bar{x} & 0 & \bar{x} & 0 & \bar{z}+1 & \bar{x} \bar{z}+\bar{x} & \bar{x} \bar{z}+\bar{x} & \bar{z}+1 \\
 0 & 0 & 0 & 0 & 0 & 0 & 0 & 0 & x \bar{z} & \bar{z} & \bar{z} & x \bar{z} \\
 0 & 0 & 0 & 0 & 0 & 0 & 0 & \bar{y} & 0 & 0 & 0 & 0 \\
 0 & 0 & 0 & z \bar{x} & z \bar{x} & 0 & z \bar{x}+\bar{x} & 0 & 0 & z \bar{x}+\bar{x} & z \bar{x}+\bar{x} & 0 \\
 0 & 0 & 0 & 0 & 1 & 1 & 0 & z \bar{y}+\bar{y} & 0 & 0 & z+1 & z+1 \\
 0 & 0 & 1 & \bar{x} & \bar{x} & 1 & 0 & 0 & \bar{z}+1 & \bar{x} \bar{z}+\bar{x} & \bar{x} \bar{z}+\bar{x} & \bar{z}+1 \\
 \bar{x} \bar{z} & 0 & 0 & 0 & \bar{x} \bar{z} & \bar{x} \bar{z} & 0 & \bar{x} \bar{y} \bar{z} & 0 & 0 & 0 & \bar{x} \bar{z}+\bar{z} \\
 0 & 0 & 0 & 0 & 0 & \bar{z} & 0 & 0 & 0 & 0 & 0 & 0 \\
 0 & \bar{y} & 0 & \bar{x} \bar{y} & \bar{x} \bar{y} & 0 & \bar{x} \bar{y} & 0 & \bar{y} \bar{z}+\bar{y} & \bar{x} \bar{y} \bar{z}+\bar{x} \bar{y} & \bar{x} \bar{y} \bar{z}+\bar{x} \bar{y} & \bar{y} \bar{z}+1 \\
\end{array}
\right)
\]

\[
l=
\left(
\begin{array}{cccccccc}
 1 & 0 & 0 & 1 & 0 & 0 & 1 & z+1 \\
 0 & 0 & 0 & 1 & 0 & 0 & 0 & 0 \\
 0 & 0 & 0 & 0 & 1 & y & 0 & 1 \\
 0 & 0 & 0 & 0 & 0 & 1 & 0 & 1 \\
 0 & 0 & 1 & x+1 & 0 & z & x+1 & x+1 \\
 0 & 0 & 0 & 0 & 0 & 0 & 1 & 1 \\
 0 & 1 & 0 & y+1 & z & 0 & 1 & 1 \\
 0 & 0 & 0 & 0 & 0 & 1 & 0 & 0 \\
\end{array}
\right)
\]
The $\bar{i}$, for $i \in \{x, y, z\}$, denotes the inverse of $i$. We find:

\[
r \cdot H^T \cdot l =
\left(
\begin{array}{cccccccc}
 1 & 0 & 0 & 0 & 0 & 0 & 0 & 0 \\
 0 & 1 & 0 & 0 & 0 & 0 & 0 & 0 \\
 0 & 0 & 1 & 0 & 0 & 0 & 0 & 0 \\
 0 & 0 & 0 & y+1 & z+1 & 0 & 0 & 0 \\
 0 & 0 & 0 & 0 & x+1 & y+1 & 0 & 0 \\
 0 & 0 & 0 & x+1 & 0 & z+1 & 0 & 0 \\
 0 & 0 & 0 & 0 & 0 & 0 & 0 & 0 \\
 0 & 0 & 0 & 0 & 0 & 0 & 0 & 0 \\
 0 & 0 & 0 & 0 & 0 & 0 & 0 & 0 \\
 0 & 0 & 0 & 0 & 0 & 0 &  \bar{x}+1 & 0 \\
 0 & 0 & 0 & 0 & 0 & 0 & \bar{z}+1 & 0 \\
 0 & 0 & 0 & 0 & 0 & 0 & \bar{y}+1 & 0 \\
\end{array}
\right)
\]
The RHS is equivalent to the 3D toric code Hamiltonian, tensored with 3 ancilla qubits. The ground state degeneracy (GSD) is given by $\log_2 \text{GSD} = \text{rank} \left( \ker(H \sigma) / H \right)$. Since both $r$ and $l$ are invertible, this model exhibits the same GSD behavior as the 3D toric code. However, since $r$ and $l$ are not symplectic, we cannot simply conclude that these two models are equivalent up to a quantum circuit.

\section{The evolution of logic operator}\label{app:logicevolution}
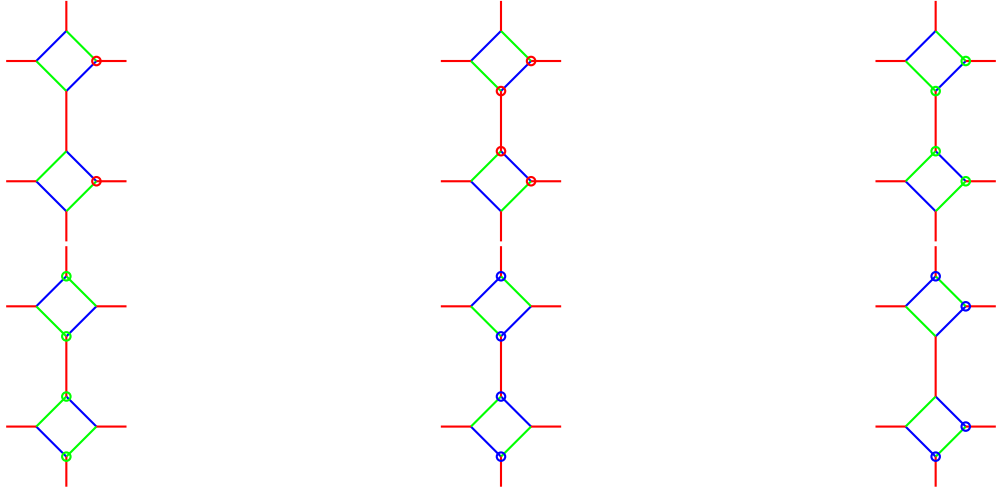
\begin{figure}[htbp]
    \centering
    
    \begin{subfigure}[b]{0.3\textwidth}
        \centering
        \begin{tikzpicture}[scale=0.4]
            \draw[thick, blue] (-1,2) -- (0,3);
            \draw[thick, blue] (0,1) -- (1,2);
            \draw[thick, green] (-1,2) -- (0,1);
            \draw[thick, green] (0,3) -- (1,2);

            \draw[thick, blue] (-1,-2) -- (0,-3);
            \draw[thick, blue] (0,-1) -- (1,-2);
            \draw[thick, green] (-1,-2) -- (0,-1);
            \draw[thick, green] (0,-3) -- (1,-2);

            \draw[thick, red] (0,-1) -- (0,1);
            \draw[thick, red] (0,4) -- (0,3);
            \draw[thick, red] (0,-3) -- (0,-4);
            \draw[thick, red] (1,2) -- (2,2);
            \draw[thick, red] (1,-2) -- (2,-2);
            \draw[thick, red] (-1,2) -- (-2,2);
            \draw[thick, red] (-1,-2) -- (-2,-2);

            \draw[thick, red] (1,2) circle (4pt);
            \draw[thick, red] (1,-2) circle (4pt);
        \end{tikzpicture}
    \end{subfigure}
    \hfill
    \begin{subfigure}[b]{0.3\textwidth}
        \centering
        \begin{tikzpicture}[scale=0.4]
            \draw[thick, blue] (-1,2) -- (0,3);
            \draw[thick, blue] (0,1) -- (1,2);
            \draw[thick, green] (-1,2) -- (0,1);
            \draw[thick, green] (0,3) -- (1,2);

            \draw[thick, blue] (-1,-2) -- (0,-3);
            \draw[thick, blue] (0,-1) -- (1,-2);
            \draw[thick, green] (-1,-2) -- (0,-1);
            \draw[thick, green] (0,-3) -- (1,-2);

            \draw[thick, red] (0,-1) -- (0,1);
            \draw[thick, red] (0,4) -- (0,3);
            \draw[thick, red] (0,-3) -- (0,-4);
            \draw[thick, red] (1,2) -- (2,2);
            \draw[thick, red] (1,-2) -- (2,-2);
            \draw[thick, red] (-1,2) -- (-2,2);
            \draw[thick, red] (-1,-2) -- (-2,-2);

            \draw[thick, red] (1,2) circle (4pt);
            \draw[thick, red] (1,-2) circle (4pt);
            \draw[thick, red] (0,1) circle (4pt);
            \draw[thick, red] (0,-1) circle (4pt);
        \end{tikzpicture}
    \end{subfigure}
    \hfill
    \begin{subfigure}[b]{0.3\textwidth}
        \centering
        \begin{tikzpicture}[scale=0.4]
            \draw[thick, blue] (-1,2) -- (0,3);
            \draw[thick, blue] (0,1) -- (1,2);
            \draw[thick, green] (-1,2) -- (0,1);
            \draw[thick, green] (0,3) -- (1,2);

            \draw[thick, blue] (-1,-2) -- (0,-3);
            \draw[thick, blue] (0,-1) -- (1,-2);
            \draw[thick, green] (-1,-2) -- (0,-1);
            \draw[thick, green] (0,-3) -- (1,-2);

            \draw[thick, red] (0,-1) -- (0,1);
            \draw[thick, red] (0,4) -- (0,3);
            \draw[thick, red] (0,-3) -- (0,-4);
            \draw[thick, red] (1,2) -- (2,2);
            \draw[thick, red] (1,-2) -- (2,-2);
            \draw[thick, red] (-1,2) -- (-2,2);
            \draw[thick, red] (-1,-2) -- (-2,-2);

            \draw[thick, green] (1,2) circle (4pt);
            \draw[thick, green] (1,-2) circle (4pt);
            \draw[thick, green] (0,1) circle (4pt);
            \draw[thick, green] (0,-1) circle (4pt);
        \end{tikzpicture}
    \end{subfigure}

    \begin{subfigure}[b]{0.3\textwidth}
        \centering
        \begin{tikzpicture}[scale=0.4]
            \draw[thick, blue] (-1,2) -- (0,3);
            \draw[thick, blue] (0,1) -- (1,2);
            \draw[thick, green] (-1,2) -- (0,1);
            \draw[thick, green] (0,3) -- (1,2);

            \draw[thick, blue] (-1,-2) -- (0,-3);
            \draw[thick, blue] (0,-1) -- (1,-2);
            \draw[thick, green] (-1,-2) -- (0,-1);
            \draw[thick, green] (0,-3) -- (1,-2);

            \draw[thick, red] (0,-1) -- (0,1);
            \draw[thick, red] (0,4) -- (0,3);
            \draw[thick, red] (0,-3) -- (0,-4);
            \draw[thick, red] (1,2) -- (2,2);
            \draw[thick, red] (1,-2) -- (2,-2);
            \draw[thick, red] (-1,2) -- (-2,2);
            \draw[thick, red] (-1,-2) -- (-2,-2);

            \draw[thick, green] (0,3) circle (4pt);
            \draw[thick, green] (0,-3) circle (4pt);
            \draw[thick, green] (0,1) circle (4pt);
            \draw[thick, green] (0,-1) circle (4pt);
        \end{tikzpicture}
    \end{subfigure}
    \hfill
    \begin{subfigure}[b]{0.3\textwidth}
        \centering
        \begin{tikzpicture}[scale=0.4]
            \draw[thick, blue] (-1,2) -- (0,3);
            \draw[thick, blue] (0,1) -- (1,2);
            \draw[thick, green] (-1,2) -- (0,1);
            \draw[thick, green] (0,3) -- (1,2);

            \draw[thick, blue] (-1,-2) -- (0,-3);
            \draw[thick, blue] (0,-1) -- (1,-2);
            \draw[thick, green] (-1,-2) -- (0,-1);
            \draw[thick, green] (0,-3) -- (1,-2);

            \draw[thick, red] (0,-1) -- (0,1);
            \draw[thick, red] (0,4) -- (0,3);
            \draw[thick, red] (0,-3) -- (0,-4);
            \draw[thick, red] (1,2) -- (2,2);
            \draw[thick, red] (1,-2) -- (2,-2);
            \draw[thick, red] (-1,2) -- (-2,2);
            \draw[thick, red] (-1,-2) -- (-2,-2);

            \draw[thick, blue] (0,3) circle (4pt);
            \draw[thick, blue] (0,-3) circle (4pt);
            \draw[thick, blue] (0,1) circle (4pt);
            \draw[thick, blue] (0,-1) circle (4pt);
        \end{tikzpicture}
    \end{subfigure}
    \hfill
    \begin{subfigure}[b]{0.3\textwidth}
        \centering
        \begin{tikzpicture}[scale=0.4]
            \draw[thick, blue] (-1,2) -- (0,3);
            \draw[thick, blue] (0,1) -- (1,2);
            \draw[thick, green] (-1,2) -- (0,1);
            \draw[thick, green] (0,3) -- (1,2);

            \draw[thick, blue] (-1,-2) -- (0,-3);
            \draw[thick, blue] (0,-1) -- (1,-2);
            \draw[thick, green] (-1,-2) -- (0,-1);
            \draw[thick, green] (0,-3) -- (1,-2);

            \draw[thick, red] (0,-1) -- (0,1);
            \draw[thick, red] (0,4) -- (0,3);
            \draw[thick, red] (0,-3) -- (0,-4);
            \draw[thick, red] (1,2) -- (2,2);
            \draw[thick, red] (1,-2) -- (2,-2);
            \draw[thick, red] (-1,2) -- (-2,2);
            \draw[thick, red] (-1,-2) -- (-2,-2);

            \draw[thick, blue] (0,3) circle (4pt);
            \draw[thick, blue] (0,-3) circle (4pt);
            \draw[thick, blue] (1,2) circle (4pt);
            \draw[thick, blue] (1,-2) circle (4pt);
        \end{tikzpicture}
    \end{subfigure}
    \caption{The evolution of the logical operator in a measurement routine. Each subfigure represents the current logical operator at each measurement round. From top-left to bottom-right, red, blue, and green checks are measured respectively. Each small (red/blue/green) circle represents a $Z/Y/X$ Pauli operator at the marked position, and the logical operator is the tensor product of the operators on the circles. It is shown that the logical operator changes its type after 3 steps of measurement and evolves back after 6 steps. The entire process requires consecutive closed spin chains connected by red checks.}
    \label{fig:evolutionoflogic}
\end{figure}

In this appendix, we will present, in Figure~\ref{fig:evolutionoflogic}, the evolution of the logical operator of the 2D toric code to demonstrate the necessary requirements to maintain the logical information of the 3D Floquet code, by doubling the measurement routine as first discussed in \cite{dua_engineering_2024}.

Where the green/blue/red segments represent the corresponding checks, and the blue/green/red circles represent the $Y/X/Z$ Pauli operators acting on the qubits at the marked vertices, forming the instantaneous logical operator. The evolution itself is not new, but it is important to highlight that this evolution can be achieved within a series of closed spin chains coupled through red checks. As mentioned earlier, the division of red and black checks is hand-chosen. Therefore, when the lattice satisfies the requirements in Section~\ref{sec:3d construction}, we can select three series of non-overlapping plaquettes along three homotopically nontrivial loops on the lattice, which is sufficient to ensure that three line logical operators of the instantaneous 3D toric code phase survive.

\section{Properties of higher dimensional $X$-cube code} \label{app:X}
As proposed in Section~\ref{subsec:construction on general lattices}, we are able to construct the Floquet code with an instantaneous phase being the higher-dimensional $X$-cube code on locally hyper-cubic lattices. In an $n$-dimensional cubic lattice, the Hamiltonian is given by Equation~\ref{eqn:higherXcube}. The number of ground state degeneracies is given by $N_q - N_S$, where $N_q$ is the number of qubits, and $N_S$ is the number of independent generators of the stabilizer group. The conditions $I_{q+1} = 0$ and $I_q \neq 0$ ensure that there is non-trivial ground state degeneracy (GSD) and no local logical operator, confirming a topological phase. Clearly, the GSD is a polynomial in the size of the cubic lattice $L$. We address the leading order of GSD as follows: 

On an $n$-dimensional $L \times L \times \dots \times L$ cubic lattice with periodic boundary condition, the translational group is represented by $\mathbf{F}_2[x_1, x_2, \dots, x_n]$, where $x^L_i = 1$. There are $L^n$ vertices, $n \cdot L^n$ edges, and $L^n$ n-cells. Since each edge has a qubit placed on it, $N_q = n \cdot L^n$. 

For the generator $A_v^{x_i,x_j}$, notice that $A_v^{x_i,x_j} \cdot A_v^{x_i,x_k} = A_v^{x_j,x_k}$. Thus, $A_v^{x_1,x_i}$ for $i = 2, 3, \dots, n$ forms a complete set of generators. With the notation of Laurent polynomials, we have the generators:
\[
A^{i,j} = (0, \dots, 1 + \bar{x}_i, 0, \dots, 1 + \bar{x}_j, \dots, 0 \, || \, 0, 0, \dots, 0),
\]
where $1+\bar{x}_i$ and $1+\bar{x}_j$ appear at the $i$-th and $j$-th positions, respectively, and $\bar{x}_i = x^{-1}_i$. The symbol $||$ represents the division between the representation of the Pauli-$X$ and Pauli-$Z$ regions. 

As discussed above, $A^{1,i}$ for $i = 2, 3, \dots, n$ constitutes a complete set of almost independent generators. Any term in the Hamiltonian can be represented as $M(x_1, x_2, \dots, x_n)A^{1,i}$, where $M(x_1, x_2, \dots, x_n)$ is a monomial of $x_1, x_2, \dots, x_n$. Therefore, we estimate that $N_{S_A} \approx (n-1) \cdot L^n$, where $\approx$ is used because these generators are not exactly independent.

To understand this dependency, note that since the lattice has periodic boundary conditions, we define $K_i = 1 + x + x^2 + \dots + x^{L-1}$ such that $K_i \cdot (1 + x_i) = 0$ over $\mathbb{F}_2$. Thus, the constraints are represented by $M(x_1, \dots, \tilde{x}_i, \dots, \tilde{x}_j, \dots, x_n) K_i K_j \cdot A^{i,j} = 0$, where $M(x_1, \dots, \tilde{x}_i, \dots, \tilde{x}_j, \dots, x_n)$ is any monomial of variables $x_1, x_2, \dots, x_n$, excluding $x_i$ and $x_j$. This contributes $\binom{n}{2} \cdot L^{n-2}$ constraints, which we call 2-constraints since they are induced from the 2D planes. We conclude that:



\[
N_{S_A} \approx (n-1) \cdot L^n - \binom{n}{2} \cdot L^{n-2}
\]

Again, we use $\approx$ since the above 2-constraints are not independent. For example,
\[
M(x_4,x_5,\dots,x_n) x^k_3 K_1 K_2 A^{1,2} = 0 \;\;\;\;\;
M(x_4,x_5,\dots,x_n) K_3 K_1 K_2 A^{1,2} = 0
\]
For $k$ ranging from $0,1,2, \dots, L-1$, the second equation is obtained by summing $k$ up in the first equation. Similarly, we get:
\[
M(x_4,x_5,\dots,x_n) K_3 K_1 K_2 A^{1,3} = 0 \;\;\;\;\;
M(x_4,x_5,\dots,x_n) K_3 K_1 K_2 A^{2,3} = 0
\]
However,
\[
M(x_4,x_5,\dots,x_n) K_3 K_1 K_2 A^{1,2} \cdot M(x_4,x_5,\dots,x_n) K_3 K_1 K_2 A^{1,3} = M(x_4,x_5,\dots,x_n) K_3 K_1 K_2 A^{3,2}
\]
The coefficients $M(x_4,x_5,\dots,x_n) K_3 K_1 K_2$ represent positions, so they have to match to allow the product as shown above. These equations demonstrate that the 2-constraints are not independent in any 3D hyperplane. We call the redundancy of 2-constraints the 3-constraints. The number of 3-constraints is clearly of the order $L^{n-3}$.




This process can be iterated: the number of $k$-constraints is of the order of $L^{n-k}$ and is further reduced by $(k+1)$-constraints, which are of the order of $L^{n-k-1}$, over $(k+1)$-dimensional hyperplanes. If we denote the number of $k$-constraints as $\xi_k$, we obtain:

\[
N_{S_A} = (n-1) \cdot L^n - \xi_2 + \xi_3 - \dots
\]

where $\xi_2 = \binom{n}{2} L^{n-2}$. Similarly, for the $B_c$ terms, they are not independent. The generator of $B_c$ is represented as:

\[
B = (0,0,\dots,0||\Xi/(1+x_1), \Xi/(1+x_2), \dots, \Xi/(1+x_n))
\]

where $\Xi = \prod_{i=1,2,\dots,n} (1 + x_i)$. Similarly, $M(x_1,\dots,\tilde{x}_i,\dots,\tilde{x}_j,\dots,x_n) K_i K_j \cdot B = 0$, and we have almost the same structure as for $A^{i,j}$. Thus, $N_{S_B} = L^n - \xi_2 + \xi'_3 - \dots$, where $\xi'_3$ is not necessarily equal to $\xi_3$. Since the generators $B_c$ and $A_v$ are mutually independent, we get $N_S = N_{S_A} + N_{S_B}$, and:

\begin{equation}
    \log_2 \text{GSD} = N_q - N_S = 2 \cdot \binom{n}{2} L^{n-2} + \text{poly}(L, n-3)
\end{equation}

where $\text{poly}(L, n-3)$ is a polynomial in $L$ with a degree less than $n-3$. This equation clearly holds in both dimension 3 and dimension 2 (which returns to the toric code), and aligns with numerical results in dimension 4.

The behavior of excitations is quite similar to that in 3D. We have lineons that move freely along specific lines but cannot turn, and membrane operators that trap excitations at the corners of the membrane. However, a pair of lineons, commonly referred to as a dipole, exhibits the same mobility as individual lineons. In 4D, these dipoles can turn when a quadrupole of lineons is paired up. More generally, in dimension $n$, an $(n-2)$-dimensional multipole can freely move in a plane perpendicular to the dimension spanned by the multipole, suggesting a higher-dimensional tensor field description.

\end{document}